\documentclass[12pt,oneside]{article}

\usepackage{amsmath,amssymb,amsfonts,mathrsfs,undertilde,latexsym}
\usepackage{graphicx}
\usepackage{enumerate,bm}
\usepackage{dsfont,comment}
\usepackage{float,multirow}
\usepackage{mwe}
\usepackage{chngcntr}
\usepackage{epstopdf}
\usepackage{cleveref}
\counterwithout{figure}{section}
\counterwithout{table}{section}
\usepackage[authoryear]{natbib}
\usepackage{pdflscape}
\usepackage{rotating}

\usepackage{multibib}
\newcites{append}{Appendix References}

\newcommand{\GG}[1]{}

\def\1{\mathbf{1}}

\def\t{\tau}
\def\<{\langle}

\def\>{\rangle}

\def\N{\mathbf{N}}

\newtheorem{pro}{Proposition}
\newtheorem{cor}{Corollary}

\def\var{\mathrm{var}}
\def\cov{\mathrm{cov}}

\def\N{\text{N}}

\def\T{{ \mathrm{\scriptscriptstyle T} }}

\def\dif{\mathrm d}

\begin{document}

\begin{titlepage}

\begin{center}
{\bf Improved Methods for Moment Restriction Models with Marginally Incompatible Data Combination and an Application to Two-sample Instrumental Variable Estimation}

\vspace{.15in} Heng Shu \& Zhiqiang Tan\footnotemark[1]

\vspace{.05in}
\end{center}

\footnotetext[1]{{ \small Heng Shu is with JPMorgan Chase, New York, NY 10017 and Zhiqiang Tan is Professor,
Department of Statistics, Rutgers University, Piscataway, NJ 08854 (E-mail: ztan@stat.rutgers.edu).
An earlier version of this work was completed as part of the PhD thesis of Heng Shu at Rutgers University.
}}

\vspace{-.3in}
\paragraph{Abstract.} {\small Combining information from multiple samples is often needed in biomedical and economic studies, but the differences between these samples must be appropriately
taken into account in the analysis of the combined data.
We study estimation for moment restriction models with data combination from two samples
under an ignorablility-type assumption but allowing for different marginal distributions of common variables between the two samples.
Suppose that an outcome regression model and a propensity score model are specified.
By leveraging the semiparametric efficiency theory, we derive an augmented inverse probability weighted (AIPW) estimator that is locally efficient and doubly robust with respect to the
outcome regression and propensity score models.
Furthermore, we develop calibrated regression and likelihood estimators that are not only locally efficient and doubly robust, but also intrinsically efficient in
achieving smaller variances than the AIPW estimator when the propensity score model is correctly specified but the outcome regression model may be misspecified. As
an important application, we study the two-sample instrumental variable problem and derive the corresponding estimators while allowing for incompatible distributions of common variables between the two samples.
Finally, we provide a simulation study and an econometric application on public housing projects
to demonstrate the superior performance of our improved estimators.}

\paragraph{Key words and phrases.} {\small Data Combination; Double robustness; Inverse probability weighting; Intrinsic efficiency; Local efficiency; Moment restriction models; Two-sample instrument variable estimation.}

\end{titlepage}

\section{Introduction} \label{intro}

Typically, empirical studies in biomedical and social sciences involve drawing inferences regarding a population.
However, there are various situations where information need to be combined from two or more samples possibly
for different populations from the target (e.g., \citealt{ridder_moffitt2007}).
For example, a single sample may not contain all the relevant variables, or some variables in the sample may be measured with errors.
Even if all the relevant variables are collected from one sample, the sample size may be too small to achieve accurate estimation.

Suppose that two random samples are obtained: a primary sample from the target population
and an auxiliary sample from another population possibly different from the target population.
The primary sample provides the measurements of variables $(Y,U)$,
and the auxiliary sample contains measurements of variables $(X,U)$.
That is, the variable $Y$ is only available from the primary data, $X$ only from the auxiliary data, and $U$ from both data.
We distinguish two different settings: \vspace{-.05in}
\begin{itemize}\addtolength{\itemsep}{-.1in}
\item[(I)] The parameter of interest can be defined through a set of moment restrictions in $(X,U)$, without involving $Y$, under the primary population (\citealt{Chen2008}).
\item[(II)] The parameter can also be defined through moment restrictions that are
separable in $(Y,U)$ and in $(X,U)$ under the primary population as studied in \cite{Graham2015}.
\end{itemize} \vspace{-.1in}
Setting (I) is more basic than (II) because the inferential difficulty mainly lies in the lack of primary data on $(X,U)$ jointly.
On the other hand, setting (I) can be subsumed under (II) with degenerate restrictions in $(Y,U)$.
A special case of such settings is estimation of average treatment effects on the treated (ATT) (\citealt{Hahn1998}).
Identification of the parameter can be achieved provided that
the conditional distributions of $X$ given $U$ are the same under the primary and auxiliary populations.
The marginal distributions of $U$ may, however, differ between the two populations.

The foregoing setting (I), with only $(X,U)$ involved but not $Y$, is called the ``verify-out-of-sample" case in \cite{Chen2008},
because the auxiliary sample is obtained independently of the primary sample such that no individual units are linked between the two samples.
This setting differs from missing-data and causal inference problems which are studied in \cite{Robins1994} and \cite{Tan2010, Tan2011} among others
and called the ``verify-in-sample" case in \cite{Chen2008} because the auxiliary sample is a subset of the primary sample by design or by happenstance.
A particular example of the latter setting is estimation of average treatment effects in the overall population (ATE) \cite[e.g.,][]{Hahn1998, Imbens2004}.
The current setting should also be contrasted with the analysis of linked data, where common units are
linked between different samples by probabilistic record linkage (e.g., \citealt{Lahiri_Larsen-2005}).

A large body of works have been done on statistical theory and methods for estimation in moment restriction models with auxiliary data in the ``verify-out-of-sample" case,
in addition to the ``verify-in-sample" case.
The semiparametric efficiency bounds are studied by \cite{Hahn1998} for ATT estimation,
by \cite{Chen2008} for moment restriction models with only $(X,U)$ involved in setting (I),
and by \cite{Graham2015} for moment restriction models that are separable in $(Y,U)$ and $(X,U)$ in setting (II).
Moreover, asymptotically globally efficient estimators are proposed in these cases by \citet{Hahn1998}, \citet{Hirano2003}, and \citet{Chen2008} among others, using nonparametric series/sieve estimation
on the propensity score (PS) or the outcome regression (OR) function.
But the smoothness conditions typically assumed for such methods can be problematic in many practical situations with a high-dimensional vector of common variables $U$ (e.g., \citealt{Robins1997}).
Recently, \citet{Graham2015} proposed a locally efficient and doubly robust method with separable moment restrictions, using parametric PS and OR models.
But methods achieving local efficiency and double robustness alone may still suffer from large variances due to inverse probability weighting.
Such a phenomenon is known in the ``verify-in-sample" case of missing-data problems (\citealt{KS2007}),
and can be seen to motivate various recent methodological development (e.g., \citealt{Tan2010, Cao2009}).

We develop improved methods for moment restriction models with data combination and make three contributions.
First, we derive augmented inverse probability weighted (AIPW) estimators in setting (I), by using efficient influence functions as estimating functions
with the true outcome regression function and propensity score replaced by their fitted values.
The idea of constructing estimating equations from influence functions (including efficient influence functions) is widely known,
at least for missing-data problems in the ``verify-in-sample" case (e.g., \citealt{Tsiatis2006, Graham2011}). But our application of this idea to
the ``verify-out-of-sample" case seems new and reveals subtle properties associated with the fact that
the semiparametric efficient bounds vary under a nonparametric model, a correctly specified
propensity score model, or known propensity scores in the ``verify-in-sample" case, instead of staying the same
in the ``verify-out-of-sample" case (\citealt{Hahn1998, Chen2008}).

On one hand, we show that the AIPW estimator based on the efficient influence function calculated under the nonparametric model
is locally nonparametric efficient (i.e., achieves the nonparametric variance bound if
both the OR and PS models are correctly specified),
and doubly robust (i.e., remains consistent if either the OR model or the PS model is correctly specified).
This AIPW estimator is simpler and more flexible than the related estimator of \citet{Graham2015}, which is shown to be locally efficient and doubly robust only
under the restrictions that the PS model is logistic regression, the OR model is linear, and all the regressors of the OR model are included in the linear span of those of the PS model.\footnotemark[2]
On the other hand, we find that the AIPW estimator based on the efficient influence function calculated with known propensity score
is locally semiparametric efficient (i.e., achieves the semiparametric variance bound calculated under the PS model used if
both the OR and PS models are correctly specified),
but generally not doubly robust.

\footnotetext[2]{For Theorem 4 of \citet{Graham2015}, it should be added to condition (b) that each component of $t(W)$ is contained in the linear span of $r(W)$.
Some of the restrictions for the method of \citet{Graham2015} can potentially be relaxed. For example, the fitted values from a nonlinear OR model
can be included as regressors in an augmented logistic PS model similarly as used in our estimators in Section~\ref{improve_est}. But our calibrated estimators are designed to
achieve desirable properties beyond local efficiency and double robustness as seen from the subsequent discussion.}

Second, we propose in setting (I) calibrated regression and likelihood estimators which are not only locally efficient and doubly robust, but also
intrinsically efficient (i.e., asymptotically more efficient than the corresponding locally efficient and doubly robust AIPW estimator when the PS model is correctly specified but the OR model may be misspecified).
Such improved estimators have been obtained in the ``verify-in-sample" case of missing-data problems (e.g., \citealt{Tan2006, Tan2010, Tan2010-CJS, Cao2009}).
But due to the aforementioned difference between the locally nonparametric and semiparametric efficient AIPW estimators, a direct application of existing techniques
would not yield an estimator with the desired properties in the ``verify-out-of-sample" setting.
We introduce a new idea to overcome this difficulty and develop estimators of the desired properties, by working with
an augmented propensity score model which includes the fitted outcome regression functions as additional regressors.

Third, our theory and methods from setting (I) can be applied and extended to setting (II). As a concrete application, we study two-sample instrumental variable estimation and derive the improved estimators in setting (II).
The two-sample instrumental variable (TSIV) estimator (\citealt{Angrist1992})
is generally consistent only when the marginal distributions
of the common variables $U$ are the same in the two samples or, equivalently, the propensity score for selection into the samples is a constant in $U$.
The two-sample two-stage least squares (TS2SLS) estimator (\citealt{BJ1997}) is consistent if either the propensity score is a constant in $U$
or the linear regression in the first stage is correctly specified.
In contrast, our calibrated estimators are doubly robust, i.e., remain consistent if either a general OR model
or a general PS model is correctly specified. Moreover, our estimators tend to achieve smaller variances than related doubly robust AIPW estimators when the PS model is correctly specified but the OR model may be misspecified. We present a simulation study and an econometric application on public housing projects, to demonstrate the advantage of our estimators compared with existing estimators.

\section{Moment restriction models with auxiliary data} \label{def}

Throughout this section, consider setting (I) described in the Introduction, where we are interested in the estimation of parameters $\theta_0$ through the moment conditions
\begin{align} \label{TS_problem}
E^{(1)}\Phi(X,U;\theta_0)=0,
\end{align}
where $E^{(1)}()$ denotes the expectation under a {\it primary} population,
$\Phi(x,u;\theta)$ is a $k\times 1$ vector of known functions, and $\theta_0$ is a $k \times 1$ vector of unknown parameters.
Suppose that $\{(X_i, U_i): i=1,\ldots,n_1\}$ are defined on an i.i.d.~sample of size $n_1$ from the primary population,
but $(X_1, \ldots, X_{n_1})$ are missing and only $(U_1, \ldots, U_{n_1})$ are observed.
For a remedy, suppose that additional data $\{(X_i, U_i): i=n_1+1,\ldots,n_1+n_0\}$ are obtained on an i.i.d.~sample of size $n_0$ from an {\it auxiliary} population,
possibly different from the primary population.
To draw valid inference about $\theta_0$, we need to combine the data $(U_1,\ldots, U_{n_1})$  from the primary population and
the data $\{(X_i, U_i): i=n_1+1,\ldots,n_1+n_0\}$ from the auxiliary population.

For technical convenience, we make the following assumption:\vspace{-.05in}
\begin{itemize}
\item[(A1)] The sample sizes $(n_1, n_0)$ are determined from binomial sampling:
the combined set of $n=n_1+n_0$ units are independently drawn from either the primary or the auxiliary population with a fixed probability $q\in (0,1)$.
As a result, $n_0/n_1$ converges in probability to a finite constant in $(0,\infty)$ as $n\to\infty$.
\end{itemize}\vspace{-.05in}
With some additional work (not pursued here), it is possible to adapt our methods and results to other sampling schemes with non-random $(n_1, n_0)$.
Under Assumption (A1), the combined set of variables $\{(T_i, U_i, X_i): i=1,\ldots,n\}$ are i.i.d.~realizations from a mixture distribution $P$,
where $T_i$ is an indicator variable, equal to either 1 or 0 if the $i$th unit is drawn from the primary or the auxiliary population.
The combined set of observed data are \vspace{-.1in}
$$
\{(T_i,U_i,(1-T_i)X_i): i=1, \ldots, n\}.
$$
The moment conditions (\ref{TS_problem}) can be represented as \vspace{-.1in}
\begin{align} \label{combine_problem}
E\left\{ \Phi(X,U;\theta_0) | T=1 \right\}=0 ,
\end{align}
where $E()$ denotes the expectation with respect to the mixture distribution $P$.
This setup is exactly the ``verify-out-of-sample'' case in \cite{Chen2008}.
Because $(X,U)$ are not jointly observed given $T=1$ (primary population), we need borrow information from $(X,U)$ jointly observed given $T=0$ (auxiliary population).

To achieve identification of $\theta_0$ by information-borrowing from the auxiliary data, there are two basic assumptions needed (e.g., \citealt{Chen2008}).
The first assumption is that the conditional distributions of $X$ given $U$ are the same under the primary and auxiliary populations, that is,
\begin{itemize} 
\item[(A2)] $X$ and $T$ are conditionally independent given $U$.
\end{itemize}
Assumption (A2) is similar to the unconfoundedness for controls for identification of ATT (e.g., \citealt{Imbens2004}).
The marginal distributions of $U$ are, however, allowed to differ between the primary and auxiliary populations.
The second assumption is that
the support of the common variable $U$ in the primary population is contained within that in the auxiliary population, that is,
\begin{itemize}  
\item[(A3)] $0\leq P(T=1|U=u)<1  \mbox{ for all } u$.
\end{itemize}
Assumption (A3) allows that $P(T=1|U=u)$ is 0 for some values $u$, i.e.,
subjects with certain $U$-values will always be in the auxiliary population, because information from those subjects is not needed for inference about $\theta_0$ in the primary population.
Nevertheless, Assumption (A3) will be strengthened in our asymptotic theory such that $0< P(T=1|U=u) \le 1-\epsilon$ for all $u$,
where $\epsilon> 0$ is a constant. See Condition (C5) and the associated discussion preceding Proposition~\ref{prop2}.

\subsection{Modeling approaches for estimation}\label{basic_approach}

There are two types of working/assisting models typically postulated for estimation of $\theta_0$ in (\ref{combine_problem}),
focusing on either the relationship between $X$ and $U$ or between $T$ and $U$,
similarly to those for estimation in missing-data problems (e.g., \citealt{KS2007}; \citealt{Tan2006, Tan2010}). The two approaches roughly correspond to
conditional expectation projection or inverse probability weighting in \cite{Chen2008}.

The first approach is to build a (parametric) regression model for the outcome regression (OR) function,
$\psi_\theta(U) =E\{\Phi(X,U;\theta)|U\}$, such that for any value $\theta$ ,
\begin{align}\label{general_OR_set}
E  \big\{ \Phi(X,U;\theta)\big|U \big\} &= \psi_\theta(U;\alpha),
\end{align}
where $\psi_\theta(u;\alpha)$ is a vector of known functions and $\alpha$ is a vector of unknown parameters.
In general, this model can be derived from a conditional density (not just mean) model of $X$ given $U$, $p(x|U;\alpha)$,
by the relationship
$\psi_\theta( U; \alpha) = \int \Phi(x, U; \theta) p(x|U; \alpha)\,\dif x$.
In special cases as discussed in Section \ref{sec:model-based}, model (\ref{general_OR_set}) can be directly induced from a conditional mean model of $X$ given $U$.
Let $\hat\alpha$ be an estimator of $\alpha$ from the auxiliary sample (i.e., $T=0$), and denote by $\hat{\psi}_\theta (U) = \psi_\theta (U;\hat{\alpha})$ the fitted outcome regression function.
Define $\hat{\theta}_{\text{\scriptsize OR}}$ as an estimator of $\theta_0$ that solves the equation
\begin{align}
\sum_{i=1}^n T_i \,\hat \psi_\theta(U_i)=0 . \label{OR_est}
\end{align}
If OR model (\ref{general_OR_set}) is correctly specified for each possible value $\theta$ (not just the true $\theta_0$), for example,
a conditional density model $p(x|U;\alpha)$ is correctly specified, then $\hat{\theta}_{\text{\scriptsize OR}}$ is a consistent estimator of $\theta_0$ under standard regularity conditions.
See Conditions (C1), (C4), and (C6) in Supplementary Material.

The other approach is to build a (parametric) regression model for the propensity score (PS), $\pi(U) =P(T=1|U)$, such that (\citealt{Rosenbaum1983})
\begin{align}
\label{TS:PS}
P( T=1| U)= \pi(U;\gamma)=\Pi\{\gamma^{\T} f(U)\},
\end{align}
where $\Pi(\cdot)$ is an inverse link function, $f(u)$ is a vector of known functions \emph{including 1}, and $\gamma$ is a vector of unknown parameters.
The score function of $\gamma$ is:
$$
S_{\gamma}(T,U)=\left\{ \frac{T}{\pi(U;\gamma)}-\frac{1-T}{1-\pi(U;\gamma)}\right\} \frac{\partial\pi(U;\gamma)}{\partial\gamma}.
$$
Typically, a logistic regression model is used:
\begin{align*} 
\pi(U; \gamma)= \big[1 +\exp\{ - \gamma^{\T} f(U)\} \big]^{-1}.
\end{align*}
Denote by $\hat\gamma$ the maximum likelihood estimator (MLE) of $\gamma$ that solves $ \tilde E \{ S_{\gamma}(T,U) \}  = 0$, which in the case of logistic regression reduces to
\begin{align}
\tilde{E} \left[ \big\{ T-\pi(U;\gamma)\big\} f(U) \right]=0 ,\label{ps_score}
\end{align}
where $\tilde{E}(\cdot)$ denotes the sample average over the merged sample.
For convenience, write the fitted propensity score as $\hat\pi(U) = \pi(U;\hat\gamma)$.
Similarly as inverse probability weighting (IPW) for the estimation of ATT (e.g., \citealt{Imbens2004}), an IPW estimator $\hat{\theta}_{\text{\scriptsize IPW}}$ for $\theta_0$ is defined as a solution to
\begin{align} \label{PS_est}
\tilde{E}\left\{\frac{1-T }{1-\hat{\pi}(U)}\hat{\pi}(U)\Phi(X,U;\theta)\right\}=0 .
\end{align}
If PS model (\ref{TS:PS}) is correctly specified, then $\hat{\theta}_{\text{\scriptsize IPW}}$ is consistent under standard regularity conditions. See Conditions (C2), (C4), (C5), and (C7) in Supplementary Material.
However, because the fitted propensity score $\hat\pi(U)$ is used for inverse weighting in Eq.~(\ref{PS_est}), $\hat{\theta}_{\text{\scriptsize IPW}}$ can be very sensitive to possible misspecification of model (\ref{TS:PS}).

\subsection{AIPW estimators}\label{variance_bound}

As discussed in Section \ref{basic_approach}, the consistency of $\hat{\theta}_{\text{\scriptsize OR}}$ depends on
the correct specification of OR model (\ref{general_OR_set}), and the consistency of $\hat{\theta}_{\text{\scriptsize IPW}}$ depends on the correct specification of PS model (\ref{TS:PS}).
We exploit semiparametric theory to derive locally efficient and doubly robust estimators of $\theta_0$ in the form of augmented IPW (AIPW),
using both OR model (\ref{general_OR_set}) and PS model (\ref{TS:PS}).
Understanding of these estimators will be important for our development of improved estimation in Section~\ref{improve_est}.

In Supplementary Material, Proposition~\ref{prop1} restates the semiparametric efficiency results from \cite{Chen2008} for estimation of $\theta_0$ under (\ref{combine_problem}) in three settings: \vspace{-.05in}
\begin{itemize} \addtolength{\itemsep}{-.1in}
\item[(i)] the propensity score $\pi(U)$ is unknown with no parametric restriction;

\item[(ii)] the propensity score $\pi(U)$ is assumed to belong to a parametric family $\pi(U;\gamma)$;

\item[(iii)] the propensity score $\pi(U)$ is known.
\end{itemize}
The efficient influence functions are denoted by $\varphi_{\text{\scriptsize NP}}$, $\varphi_{\text{\scriptsize SP}}$, and $\varphi_{\text{\scriptsize SP*}}$,
and their variances (i.e., semiparametric efficiency bounds) are denoted by $V_{\text{\scriptsize NP}}$, $V_{\text{\scriptsize SP}}$, and $ V_{\text{\scriptsize SP*}}$.
It holds, in general with strict inequalities, that $V_{\text{\scriptsize NP}} \ge V_{\text{\scriptsize SP}} \ge V_{\text{\scriptsize SP*}}$, which
is in contrast with other missing-data problems such as Robins et al.~(1994) and the ``verify-in-sample" case in Chen et al.~(2008), where $V_{\text{\scriptsize NP}}=V_{\text{\scriptsize SP}} = V_{\text{\scriptsize SP*}}$.

Two estimator of $\theta_0$ can be derived by directly taking the efficient influence functions as estimating functions,
with the unknown true functions $\psi_\theta(U)$ and $\pi(U)$ replaced by
the fitted values $\hat{\psi}_\theta(U)$ and $\hat{\pi}(U)$.
The first estimator, denoted by  $\hat{\theta}_{\text{\scriptsize NP}}$, is based on $\varphi_{\text{\scriptsize NP}}$ and defined as a solution to
\begin{align} \label{est_NP}
\tilde{E}\left[ \frac{1-T}{1-\hat{\pi}(U)}\hat{\pi}(U)\big\{ \Phi(X,U;\theta)-\hat{\psi}_\theta(U)\big\} +
T \hat{\psi}_\theta(U)\right] =0 .
\end{align}
The second estimator, denoted by  $\hat{\theta}_{\text{\scriptsize SP}}$, is based on $\varphi_{\text{\scriptsize SP}}$ or equivalently based on $\varphi_{\text{\scriptsize SP*}}$ and defined as a solution to
\begin{align} \label{est_SP}
\tilde{E}\left[ \frac{1-T}{1-\hat{\pi}(U)}\hat{\pi}(U)\big\{ \Phi(X,U;\theta)-\hat{\psi}_\theta(U)\big\} +
\hat{\pi}(U)\hat{\psi}_\theta(U)\right] =0 .
\end{align}
Proposition \ref{prop2} shows that both estimators possess local efficiency but of different types,\footnotemark[3] and only  $\hat{\theta}_{\text{\scriptsize NP}}$ is doubly robust.
For clarity, the semiparametric efficiency bound $V_{\text{\scriptsize NP}}$ under the nonparametric PS model is hereafter called the \emph{nonparametric efficiency bound}.
See, for example, \citet{Newey1990}, \citet{Robins2001}, and \citet{Tsiatis2006} for general discussions on local efficiency and double robustness.

\footnotetext[3]{Local efficiency means attaining the efficiency bound under a semiparametric model when some additional modeling restrictions (not part of the model) are satisfied.
Thus there are different types of local efficiency, depending on what model and additional restrictions are involved.}

We briefly describe regularity conditions for the asymptotic results below. See Appendix~\ref{tech_details} in Supplementary Material for details.
To match the Supplementary Material, the numbering of the conditions is not consecutive here. \vspace{-.05in}
\begin{itemize}\addtolength{\itemsep}{-.1in}
\item[(C1)] For a constant $\alpha^*$, it holds that $ \hat \alpha = \alpha^* + O_p(n^{-1/2})$. If OR model (\ref{general_OR_set}) is correctly specified, then
$\psi_\theta(U) = E\{\Phi(X,U;\theta)| U\} = \psi_\theta(U;\alpha^*)$.

\item[(C2)] For a constant $\gamma^*$, it holds that $ \hat \gamma = \gamma^* + O_p(n^{-1/2})$. If PS model (\ref{TS:PS}) is correctly specified, then
$\pi(U) = P(T=1|U) = \pi(U;\gamma^*)$.

\item[(C4)] The vector of estimating functions $T \Phi(X,U;\theta)$ satisfies regularity conditions to ensure $n^{-1/2}$-convergence of
the estimator of $\theta_0$ that solves $0 = \tilde E\{T \Phi(X,U;\theta)\}$ if, hypothetically, $(X,U)$ were jointly observed given $T=1$.

\item[(C5)] There exists a constant $\epsilon>0$ such that $0<\pi(u;\gamma^*) \le 1-\epsilon$ for all $u$.
We assume that $\pi(u;\gamma^*)$ is bounded away from 1, to avoid ``irregular identification" for inverse weighting (\citealt{Khan-Tamer2010}).
Moreover, we assume that $\pi(u;\gamma^*)$ is nonzero (but possibly close to 0), to simplify technical arguments; otherwise, some components of $\gamma^*$ would be $\pm \infty$ in PS model (\ref{TS:PS}).

\setlength{\itemindent}{-.4in}

\item[] (C7)--(C8) Partial derivative matrices of $\psi_\theta(U;\alpha)$ and $\pi(U;\gamma)$ are uniformly integrable in neighborhoods of $\theta_0$, $\alpha^*$, and $\gamma^*$.
\end{itemize}

\begin{pro} \label{prop2}
In addition to Assumptions~(A1)--(A2), suppose that Conditions (C1), (C2), (C4), (C5), (C7), and (C8) are satisfied, allowing for possible model misspecification (e.g., \citealtappend{White1982}).
Then the following results hold.\vspace{-.05in}
\begin{enumerate}\addtolength{\itemsep}{-.1in}
\item[(i)] The estimator $\hat{\theta}_{\text{\scriptsize NP}}$ is doubly robust: it remains consistent when either model (\ref{general_OR_set}) or model (\ref{TS:PS}) is correctly specified.
Moreover, $\hat{\theta}_{\text{\scriptsize NP}}$ is locally nonparametric efficient: it achieves the nonparametric efficiency bound $V_{\text{\scriptsize NP}}$
when both model (\ref{general_OR_set}) and model (\ref{TS:PS}) are correctly specified.

\item[(ii)] The estimator $\hat{\theta}_{\text{\scriptsize SP}}$ is locally semiparametric efficient:  it achieves the semiparametric efficiency bound $V_{\text{\scriptsize SP}}$ when both model (\ref{general_OR_set}) and model (\ref{TS:PS}) are correctly specified. But $\hat{\theta}_{\text{\scriptsize SP}}$ is, generally, not doubly robust.
\end{enumerate}
\end{pro}

For both estimators $\hat{\theta}_{\text{\scriptsize NP}}$ and $\hat{\theta}_{\text{\scriptsize SP}}$,
the estimating equations (\ref{est_NP}) and (\ref{est_SP}) can be expressed in the following AIPW form
with the choice $h(U)=\hat{\psi}_\theta(U)$ or $h(U)=\hat{\pi}(U)\hat{\psi}_\theta(U)$ respectively:
\begin{align}
\tilde{E}\left[ \frac{1-T}{1-\hat{\pi}(U)}\hat{\pi}(U)\Phi(X,U;\theta)-\left\{ \frac{1-T}{1-\hat{\pi}(U)}-1\right\} h(U)\right] =0 .  \label{AIPW_eq}
\end{align}
Setting $h(U)\equiv0$ leads to the IPW estimator $\hat{\theta}_{\text{\scriptsize IPW}}$.
By local semiparametric efficiency in Proposition \ref{prop2}(ii), $\hat{\theta}_{\text{\scriptsize SP}}$ achieves the minimum asymptotic variance among all regular estimators under PS model (\ref{TS:PS}), including
AIPW estimators like $\hat{\theta}_{\text{\scriptsize NP}}$ as solutions to (\ref{AIPW_eq}) over possible choices of $h(U)$, when both PS model (\ref{TS:PS})
and OR model (\ref{general_OR_set}) are correctly specified. However, $\hat{\theta}_{\text{\scriptsize SP}}$ is not doubly robust,
and $\hat{\theta}_{\text{\scriptsize NP}}$ is doubly robust.
This situation should be contrasted with other missing-data problems such as the ``verify-in-sample" case where nonparametric and semiparametric efficiency bounds are the same,
and there exists an AIPW estimator that is locally nonparametric and semiparametric efficient and doubly robust simultaneously (e.g., \citealt{Robins1994, Tan2006, Tan2010}).
These differences present new challenges in our development of improved estimation; see the discussion after Proposition~\ref{prop-reg}.

\subsection{Improved estimation} \label{improve_est}

We develop improved estimators of $\theta_0$ under moment conditions (\ref{combine_problem}) which are not only doubly robust and locally nonparametric efficient, but also intrinsically efficient: as long as PS model (\ref{TS:PS}) is correctly specified, these estimators will attain the smallest asymptotic variance among a class of AIPW estimators including $\hat{\theta}_{\text{\scriptsize NP}}$ but with $\hat{\pi}(U)$ replaced by the fitted value from an augmented propensity score model as defined later in (\ref{TS:augPS}).
The new estimators are then similar to $\hat{\theta}_{\text{\scriptsize NP}}$ in achieving local nonparametric efficiency and double robustness, but
often achieve smaller variances than $\hat{\theta}_{\text{\scriptsize NP}}$ when
PS model (\ref{TS:PS}) is correctly specified but the OR model is misspecified.

\subsubsection{Calibrated regression estimator} \label{sec:cal-reg}

We derive regression estimators for $\theta_0$, similar to the regression estimators of ATE in \citet{Tan2006}, but with an important new idea as follows.
For simplicity, assume that PS model (\ref{TS:PS}) is logistic regression.
With additional technical complexity, the approach can be extended when PS model (\ref{TS:PS}) is non-logistic regression similarly as discussed in \cite{Shu2015}.
Consider an augmented PS model
\begin{align}
& P(T=1| U) = \pi_{\text{\scriptsize aug}}(U; \gamma,\delta, \hat\alpha ) \nonumber \\
& = \mbox{expit} \left\{ \gamma^\T f(U) + \delta^\T\hat{\psi}_\theta(U) \right\}, \label{TS:augPS}
\end{align}
where $\mbox{expit}(c)=\{1+\exp(-c)\}^{-1}$
and $\delta$ is a $k\times 1$ vector of unknown coefficients for additional regressors $\hat{\psi}_\theta(U) = \psi_\theta(U;\hat\alpha)$.
Let $(\tilde\gamma, \tilde\delta)$ be the MLE of $(\gamma,\delta)$ and $\tilde\pi(U) = \pi_{\text{\scriptsize aug}}(U;\tilde\gamma,\tilde\delta, \hat\alpha)$,
depending on $\theta$ through $\hat\psi_\theta(U)$. This dependency on $\theta$ is suppressed for convenience in the notation.
A consequence of including the additional regressors is that, by Eq.~(\ref{ps_score}), we have the two equations,
\begin{align}
& \tilde E\big[  \{T-\tilde\pi(U)\} f(U)\big]=0,  \label{augPS_eq} \\
& \tilde E\big[ \{T- \tilde\pi(U) \}\hat{\psi}_\theta(U)  \big] = 0 .  \label{augPS_eq2}
\end{align}
For augmented PS model (\ref{TS:augPS}), there may be linear dependency in the variables $\{f(U)$, $\hat{\psi}_\theta(U)\}$. In this case,
the regressors should be redefined to remove redundancy.

We define the regression estimator $\tilde{\theta}_{\text{\scriptsize reg}}$ as a solution to
\begin{align}\label{est_reg}
\tilde{E} \{ \tilde{\tau}_{\text{\scriptsize reg}}(\theta) \}=0
\end{align}
with $\tilde\tau_{\text{\scriptsize reg}}(\theta)=\tilde{\tau}_{\text{\scriptsize init}}(\theta) -\tilde{\beta }^\T(\theta)  \tilde{\xi }$ and
$\tilde{\beta }(\theta)= \tilde{E}^{-1}( \tilde{\xi } \tilde{\zeta }^\T )\tilde{E}\{ \tilde{\xi } \tilde{\tau}_{\text{\scriptsize init}}^{\T}(\theta) \}$,
where
\begin{align*}
& \tilde{\tau}_{\text{\scriptsize init}}(\theta) = \frac{1-T}{1-\tilde{\pi} (U)}\tilde{\pi} (U)\Phi(X,U;\theta) ,\\
& \tilde{\xi} =\left\{ \frac{1-T}{1-\tilde{\pi} (U)}-1\right\} \frac{\tilde{h} (U)}{\tilde{\pi} (U)}, \quad
 \tilde{\zeta }=\frac{1-T}{1-\tilde{\pi} (U)}\frac{\tilde{h} (U)}{\tilde{\pi} (U)},
\end{align*}
and $\tilde{h }(U )=\{\tilde{h }_1^\T (U ), \tilde{h }_2^\T (U )\}^\T $ are defined as follows,
\begin{align*}
\tilde{h }_1( U ) &= \tilde{\pi} (U) \tilde v(U), \quad \tilde v(U) = \left\{ \tilde{\pi} ( U ),\tilde{\pi} ( U )\hat{\psi}_\theta^\T( U)\right\}^\T, \\
\tilde{h }_2(U )&=\tilde{\pi} (U) \{1-\tilde{\pi} (U)\} \,\left\{f^\T(U),\hat{\psi}_\theta^\T(U) \right\}^\T .
\end{align*}
The dependency of $\tilde h$, $\tilde\xi$, and $\tilde\zeta$ on $\theta$ through $\hat\psi_\theta(U)$ is suppressed in the notation.
To compute $\tilde{\theta}_{\text{\scriptsize reg}}$, the equations (\ref{augPS_eq})--(\ref{augPS_eq2}) and (\ref{est_reg})
can be solved jointly by alternating Newton-Raphson iterations to update $(\gamma,\delta)$ and $\theta$, as in \cite{Tan2010-CJS}.
The computation can be simplified in special cases, as discussed in Section \ref{sec:reg-lik}.

The  variables in $\tilde h(U)$ are included to achieve different properties. First, $\tilde \pi(U)$ is included in $\tilde v(U)$ to ensure efficiency gains over the ratio estimator.
Second, $\tilde{\pi}( U )\hat{\psi}_\theta( U)$ is included
in $\tilde v(U)$ to achieve double robustness and local nonparametric efficiency, as later seen from Eq.~(\ref{reason_DR}).
Finally, $\tilde h_2(U)$ is included to account for the variation of $(\tilde\gamma,\tilde\delta)$ to achieve intrinsic efficiency as described in Proposition \ref{prop-reg} below.
The variables in $\tilde\xi$ corresponding to $\tilde h_2(U)$ are exactly the scores $\{T-\tilde\pi(U)\}\{f^\T(U), \hat \psi_\theta^\T(U)\}^\T$ for the augmented PS model (\ref{TS:augPS}).
The subvector $\tilde h_2(U)$ can be removed to reduce the dimension of $\tilde h(U)$, with little sacrifice or even improvement in finite samples.


We impose the following regularity conditions in addition to those described earlier for Proposition~\ref{prop2}. See Appendix~\ref{tech_details} in Supplementary Material for details.
\begin{itemize} \addtolength{\itemsep}{-.1in}
\item[(C3)] For $\theta$ in a neighborhood $N_0$ of $\theta_0$ and some constants $\{\gamma^\dag(\theta),\delta^*(\theta)\}$, it holds that $(\tilde\gamma,\tilde\delta) = \{\gamma^\dag(\theta),\delta^*(\theta)\} + O_p(n^{-1/2})$. If PS model (\ref{TS:PS}) is correctly specified, then $\pi(U) = P(T=1|U) = \pi_{\text{\scriptsize aug}}(U;\gamma^*,0, \alpha^*)$.

\item[(C6)] There exists a constant $\epsilon>0$ such that $0<\pi_{\text{\scriptsize aug}}\{U;\gamma^\dag(\theta), \delta^*(\theta), \alpha^*\} \le 1-\epsilon$ for
all $u$ and $\theta \in N_0$.

\item[(C9)] Partial derivative matrices of $\pi_{\text{\scriptsize aug}}(U;\gamma, \delta, \alpha)$ are uniformly integrable in neighborhoods of $\{\gamma^\dag(\theta), \delta^*(\theta), \alpha^*\}$ for $\theta\in N_0$.
\end{itemize}
Assumption (C6) is similar to (C5), whereas (C9) is similar to (C8). In particular, if PS model (\ref{TS:PS}) is correctly specified, then (C6) is equivalent to (C5).
If PS model (\ref{TS:PS}) is misspecified, then (C6) requires that the limit propensity score under the augmented PS model (\ref{TS:augPS}) is bounded away from 1 for $\theta\in N_0$.

\begin{pro} \label{prop-reg}
Suppose that Assumptions~(A1)--(A2) and Conditions (C1)--(C9) are satisifed, and PS model (\ref{TS:PS}) is logistic regression.
Then the following results hold.
\begin{enumerate}\addtolength{\itemsep}{-.1in}
\item[(i)] $\tilde\theta_{\text{\scriptsize reg}}$ is doubly robust:
it remains consistent when either model (\ref{general_OR_set}) or model (\ref{TS:PS}) is correctly specified.

\item[(ii)] $\tilde\theta_{\text{\scriptsize reg}}$ is locally nonparametric efficient:
it achieves the nonparametric efficiency bound  $V_{\text{\scriptsize NP}}$  when both model (\ref{general_OR_set}) and model (\ref{TS:PS}) are correctly specified.

\item[(iii)] $\tilde\theta_{\text{\scriptsize reg}}$ is intrinsically efficient:
if model (\ref{TS:PS}) is correctly specified, then it achieves the lowest asymptotic variance among the class of estimators of $\theta_0$ that are solutions to $k$ estimating equations of the form
\begin{align} \label{theta_class}
\tilde{E}\left\{ \tilde{\tau}_{\text{\scriptsize init}}(\theta)-b^\T\tilde{\xi}\right\} =0,
\end{align}
where $b$ is a $\text{dim}(h)\times k$ matrix of constants.
\end{enumerate}
\end{pro}

In the following, we provide several remarks to discuss Proposition~\ref{prop-reg}.

\vspace{.05in}
{\bf Double robustness}.\;
We explain why the use of the augmented propensity score $\tilde{\pi}(U)$ is important for $\tilde\theta_{\text{\scriptsize reg}}$ to achieve double robustness,
in addition to the fact that $\tilde{\pi}( U )\hat{\psi}_\theta( U)$ is included in $\tilde v(U)$.
If the OR model (\ref{general_OR_set}) is correctly specified, then, as shown in the proof of Proposition~\ref{prop-reg} in Supplementary Material,
$\tilde\theta_{\text{\scriptsize reg}}$ is asymptotically equivalent, up to $o_p(n^{-1/2})$, to a solution of the equation
\begin{align} \label{reason_DR}
\tilde{E}\left[ \frac{1-T}{1-\tilde{\pi}(U)}\tilde{\pi}(U) \{\Phi(X,U;\theta)-\hat{\psi}_\theta (U)\} +\tilde{\pi}(U)\hat{\psi}_\theta (U)\right]=0,
\end{align}
mainly because $\tilde{\pi}(U)\hat{\psi}_\theta (U)$ is included in $\tilde v(U)$.
By the use of the augmented PS model, Eq.~(\ref{augPS_eq2}) holds and hence Eq.~(\ref{reason_DR}) is identical to the equation
\begin{align} \label{TS:tilde_NP}
\tilde{E}\left[ \frac{1-T}{1-\tilde{\pi}(U)}\tilde{\pi}(U)\{\Phi(X,U;\theta)-\hat{\psi}_\theta (U)\} + T\hat{\psi}_\theta (U)\right]= 0,
\end{align}
which has exactly the same form as Eq.~(\ref{est_NP}) but with $\hat{\pi}(U)$ replaced by the augmented propensity score $\tilde{\pi}(U)$.
Let $\tilde{\theta}_{\text{\scriptsize NP}}$ be a solution of Eq.~(\ref{TS:tilde_NP}). Then $\tilde{\theta}_{\text{\scriptsize NP}}$ is doubly robust
similarly as $\hat{\theta}_{\text{\scriptsize NP}}$ based on $\hat{\pi}(U)$ by Proposition~\ref{prop2}.
Therefore, $\tilde{\theta}_{\text{\scriptsize reg}}$ is consistent when OR model (\ref{general_OR_set}) is correctly specified even if the PS model (\ref{TS:PS}) is misspecified.

\vspace{.05in}
{\bf Local efficiency}.\;
The asymptotic equivalence between $\tilde{\theta}_{\text{\scriptsize reg}}$ and $\tilde{\theta}_{\text{\scriptsize NP}}$ discussed above under OR model (\ref{general_OR_set}) also
implies that
when model (\ref{general_OR_set}) is correctly specified, $\tilde{\theta}_{\text{\scriptsize reg}}$ is locally nonparametric efficient, similarly as $\tilde \theta_{\text{\scriptsize NP}}$
and $\hat\theta_{\text{\scriptsize NP}}$.
It should be noted that $\tilde{\theta}_{\text{\scriptsize reg}}$ is generally not locally semiparametric efficient in terms of
PS model (\ref{TS:PS}), but locally semiparametri efficient in terms of PS model (\ref{TS:augPS}):
$\tilde{\theta}_{\text{\scriptsize reg}}$ achieves the semiparametric efficiency bounded calculated under model (\ref{TS:augPS}), not under model (\ref{TS:PS}),
when both model (\ref{general_OR_set}) and model (\ref{TS:PS}) are correctly specified. In fact, when PS model (\ref{TS:PS}) holds,
the efficiency bound $V_{\text{\scriptsize SP}}$ under model (\ref{TS:augPS}) coincides with the
nonparametric efficiency bound $V_{\text{\scriptsize NP}}$, because $\{T -\pi(U)\} \psi_\theta(U)$
can be shown to be included in the score function of model (\ref{TS:augPS}).
On the other hand, $\tilde{\theta}_{\text{\scriptsize reg}}$ with $\tilde\pi(U)$ replaced by $\hat \pi(U)$ throughout would be locally semiparametric efficient with respect to original PS model (\ref{TS:PS}), but generally
not doubly robust, similarly as $\hat{\theta}_{\text{\scriptsize SP}}$.

\vspace{.05in}
{\bf Intrinsic efficiency}.\;
The regression coefficient $\tilde \beta(\theta)$ is constructed by the approach of \citet{Tan2006}, for $\tilde\theta_{\text{\scriptsize reg}}$ to achieve intrinsic efficiency
beyond local nonparametric efficiency and double robustness. In fact, we did not apply
$\hat \beta(\theta) = \tilde{E}( \tilde{\xi }\tilde{\xi }^\T )^{-1}\tilde{E}\{ \tilde{\xi }\tilde{\tau}_{\text{\scriptsize init}}(\theta)\}$,
the classical estimator of the optimal choice $b$ in minimizing the asymptotic variance of (\ref{theta_class}).
The estimator $\hat{\theta}_{\text{\scriptsize reg}}$, which solves the equation
$\tilde{E} \{\tilde{\tau}_{\text{\scriptsize init}}(\theta) -\hat{\beta}^\T(\theta)  \tilde{\xi}\} =0$,
is asymptotically equivalent to the first order to $\tilde{\theta}_{\text{\scriptsize reg}}$
when the PS model is correctly specified.
But $\hat{\theta}_{\text{\scriptsize reg}}$, unlike $\tilde{\theta}_{\text{\scriptsize reg}}$, is generally inconsistent for $\theta_0$, when OR model is correctly specified and PS model may be misspecified.
The particular form of $\tilde\beta(\theta)$ can also be derived through empirical efficiency maximization (\citealt{Rubin2008, Tan2008})
and design-optimal regression estimation for survey calibration (\citealt{Tan2013}).
See further discussion related to calibration estimation after Proposition \ref{prop-lik}.

The advantage of achieving intrinsic efficiency can be seen as follows. Let $\tilde{\theta}_{\text{\scriptsize IPW}}$ and,
as done before, $\tilde{\theta}_{\text{\scriptsize NP}}$
be a solution to Eq.~(\ref{PS_est}) for $\hat{\theta}_{\text{\scriptsize IPW}}$ and Eq.~(\ref{est_NP}) for $\hat{\theta}_{\text{\scriptsize NP}}$ respectively,
with $\hat{\pi}(U)$ replaced by $\tilde{\pi}(U)$.
Moreover, consider an extension of the auxiliary-to-study tilting (AST) estimator in \cite{Graham2015} under our general setting
using the augmented PS model (\ref{TS:augPS}) as
mentioned in Introduction. Let $\tilde \theta_{\text{\scriptsize AST}}$ be a solution to $\tilde E \{ (1-T) \tilde w^{-1}_{\text{\scriptsize AST}}(U) \tilde \pi(U) \Phi(X,U;\theta) \}=0$,
where $\tilde w_{\text{\scriptsize AST}}(U) = 1- \mbox{expit}\{ \tilde\gamma^\t f(U) + (\tilde\delta +\tilde\chi)^\T \hat \psi_\theta (U) \}$,
$(\tilde\gamma,\tilde \delta)$ is the MLE of $(\gamma,\delta)$ for model (\ref{TS:augPS}), and $\tilde\chi$ is chosen such that
$\tilde E \big[\{ (1-T) \tilde w^{-1}_{\text{\scriptsize AST}}(U)-1\} \tilde \pi(U) \hat \psi_\theta(U) \big]=0$.
Then  $\tilde{\theta}_{\text{\scriptsize NP}}$ and $\tilde \theta_{\text{\scriptsize AST}}$ can be shown to be doubly robust and locally nonparametric efficient.

\begin{cor} \label{cor-reg}
Under the setting of Proposition~\ref{prop-reg}, if PS model (\ref{TS:PS}) is correctly specified, then the estimator $\tilde{\theta}_{\text{\scriptsize reg}}$ is asymptotically at least as efficient as
$\tilde{\theta}_{\text{\scriptsize IPW}}$, $\tilde{\theta}_{\text{\scriptsize NP}}$, and $\tilde{\theta}_{\text{\scriptsize AST}}$.
\end{cor}

The concept of intrinsic efficiency was introduced in related works on missing-data and causal inference problems \citep{Tan2006, Tan2010, Tan2010-CJS}, and is useful
for comparing various estimators that are all shown to be doubly robust and locally efficient when both OR and PS models are involved.
Roughly speaking, intrinsic efficiency indicates that an estimator achieves the smallest possible asymptotic variance among a class of AIPW-type estimators, such as (\ref{theta_class}), using the same
fitted OR function as long as the PS model is correctly specified.
It is tempting, but remains an open question, to formulate a similar property in terms of a correctly specified OR model and construct estimators with the desired property.
See \cite{Tan2007} for a discussion about different characteristics of PS and OR approaches related to DR estimation.

\subsubsection{Calibrated likelihood estimator}

A practical limitation of the regression estimator $\tilde{\theta}_{\text{\scriptsize reg}}$ is that
it may take some outlying values
due to large inverse weights $\{1-\tilde{\pi}(U)\}^{-1}$ in both terms $\tilde{\tau}_{\text{\scriptsize init}}(\theta)$ and $ \tilde{\xi }$.
In this section, we derive a likelihood estimator of $\theta$ which is doubly robust, locally nonparametrically efficient and intrinsically efficient similarly to the regression estimator $\tilde{\theta}_{\text{\scriptsize reg}}$,
but tends to be less sensitive to large weights than the regression estimator. 

We take two steps to derive a likelihood estimator achieving all the desirable properties.
First, we derive a locally nonparametric efficient, intrinsically efficient, but non-doubly robust estimator of $\theta_0$ by the approach of empirical likelihood using estimating equations (\citealt{Owen2001, Qinlawless1994})
or equivalently the approach of nonparametric likelihood  (\citealt{Tan2006,Tan2010}).
Specifically, our approach is to maximize the log empirical likelihood $\sum_{i=1}^n\log p_i$ subject to the constraints
\begin{align} \label{emp-lik-constraints}
\sum_{i=1}^n p_i=1, \quad
\sum_{i=1}^n p_i \tilde \xi_{i}= 0,
\end{align}
where $p_i$ is a nonnegative weight assigned to $(T_i, X_i, U_i)$ for $i = 1,...,n$ with $\sum_{i=1}^n p_i=1$.
Let $\{\hat p_i: i=1,\ldots,n\}$ be the weights obtained from the maximization. The empirical likelihood estimator of $\theta_0$, $\hat{\theta}_{\text{\scriptsize lik}}$, is defined as a solution to
\begin{align}\label{lik_ndr}
\sum_{i=1}^n \hat{p}_i\left\{ \frac{1-T_i}{1-\tilde{\pi}(U_i)}\tilde{\pi}(U_i)\Phi(X_i,U_i;\theta)\right\}=0,
\end{align}
where $\tilde{\pi}(U)$ is the fitted value of $P(T=1|U)$ under model (\ref{TS:augPS}) as in Section \ref{sec:cal-reg}.
In the just-identified setting with $\Phi()$ and $\theta$ of the same dimension,
this approach is equivalent to maximizing the empirical likelihood subject to (\ref{emp-lik-constraints}) and (\ref{lik_ndr}) together.
In Supplementary Material, we show that Eq.~(\ref{lik_ndr}) can also be expressed as
\begin{align}\label{lik1}
\frac{1}{n}\sum_{i=1}^n\left\{ \frac{1-T_i}{1-\omega(U_i;\hat{\lambda})}\tilde{\pi}(U_i)\Phi(X_i,U_i;\theta)\right\} =0 ,
\end{align}
where $\omega(U ;\lambda)=\tilde{\pi}(U )+\lambda^\T\tilde{h }(U )$ and $\hat{\lambda}$ is a maximizer of the function
\begin{align*}
\ell(\lambda)=\tilde{E}\Big[ T\log{\omega(U ;\lambda)}+(1-T)\log\{ 1-\omega(U ;\lambda) \} \Big] ,
\end{align*}
subject to $\omega(U_i;\lambda)>0$ if $T_i=1$ and $\omega(U_i;\lambda)<1$ if $T_i=0$ for $i=1,\ldots,n$.
Setting the gradient of $\ell(\lambda)$ to zero shows that $\hat{\lambda}$ is a solution to
\begin{eqnarray}
\tilde{E}\left[ \frac{T-\omega(U ;\lambda)}{\omega(U ;\lambda) \{1-\omega(U ;\lambda)\}} \tilde{h }(U) \right]=0 . \label{lik-score}
\end{eqnarray}

The estimator $\hat{\theta}_{\text{\scriptsize lik}}$ can be shown to be intrinsically efficient among the class of estimators (\ref{theta_class})
and locally nonparametric efficient, but generally not doubly robust.
Next we introduce the following modified likelihood estimator, to achieve double robustness but without affecting the first-order asymptotic behavior.

Partition $\tilde{h }$ as $\tilde h= (\tilde{h }_{1}^\T, \tilde{h }_2^\T)^\T$ and accordingly
$\lambda$ as $\lambda =(\lambda _{1}^\T, $ $\lambda _2^\T)^\T$.
Define $\tilde \lambda = (\tilde \lambda_{1}^\T, \hat\lambda_2^\T)^\T$, where $ \hat\lambda_2$ are obtained from $\hat\lambda$, and
$\tilde \lambda_{1}$ is a maximizer of the function
$$
\kappa(\lambda _{1})=\tilde{E}\left[ (1-T)\frac{\log\{1-\omega(U;\lambda _{1},\hat{\lambda }_2)\}-
\log\{1-\omega(U ;\hat{\lambda })\}}{\tilde{\pi}(U )}-\lambda _{1}^\T \tilde{v} (U)\right],
$$
subject to $\omega(U_i;\lambda _{1},\hat{\lambda }_2)<1 $ if $T_i=0$ for $i=1,\ldots,n$.
Setting the gradient of $\kappa(\lambda _{1})$ to 0 shows that $\tilde{\lambda }_{1}$ is a solution to
\begin{eqnarray}
\tilde{E}\left[\left\{ \frac{1-T}{1-\omega(U ;\lambda_{1},\hat{\lambda }_2)}-1\right\} \tilde{v} (U )\right]=0. \label{eq_general_lik2}
\end{eqnarray}
The resulting estimator of $\theta_0$, $\tilde{\theta}_{\text{\scriptsize lik}}$, is defined as a solution to
\begin{align} \label{eq_general_lik}
\tilde{E}\left\{\frac{1-T}{1-\omega(U;\tilde{\lambda})}\tilde{\pi}(U)\Phi(X,U;\theta)\right\}=0 ,
\end{align}
where $\theta$ is also involved in $\tilde\pi(U)$ and $\omega(U;\tilde{\lambda})$, although this dependency is suppressed in the notation.
To compute $\tilde{\theta}_{\text{\scriptsize lik}}$, the equations (\ref{augPS_eq})--(\ref{augPS_eq2}), (\ref{lik-score})--(\ref{eq_general_lik2}),
and (\ref{eq_general_lik}) can be solved by alternating Newton-Raphson iterations. See Section \ref{sec:reg-lik} for simplification in special cases.
The estimator $\tilde{\theta}_{\text{\scriptsize lik}}$ has several desirable properties as follows.\vspace{-.1in}

\begin{pro} \label{prop-lik}
Under the setting of Proposition~\ref{prop-reg}, the estimator $\tilde\theta_{\text{\scriptsize lik}}$
has the following properties. \vspace{-.1in}
\begin{enumerate}\addtolength{\itemsep}{-.1in}
\item[(i)] $\tilde\theta_{\text{\scriptsize lik}}$ is doubly robust, similarly as $\tilde\theta_{\text{\scriptsize reg}}$ in Proposition \ref{prop-reg}.

\item[(ii)] If model (\ref{TS:PS}) is correctly specified, then
$\tilde\theta_{\text{\scriptsize lik}}$ is asymptotically equivalent, to the first order, to $\tilde\theta_{\text{\scriptsize reg}}$. Hence
$\tilde\theta_{\text{\scriptsize lik}}$ is intrinsically efficient among the class (\ref{theta_class}) and locally nonparametric efficient, similarly as $\tilde\theta_{\text{\scriptsize reg}}$ in Proposition \ref{prop-reg}.
\end{enumerate}
\end{pro}

The double robustness of $\tilde{\theta}_{\text{\scriptsize lik}}$ holds mainly for two reasons. First, we have
$\tilde{E}[(1-T)\tilde{\pi}(U)\hat{\psi}_\theta(U)/\{1-\omega(U;\tilde{\lambda}_{1},\hat{\lambda }_2)\} ] =
\tilde{E}\{\tilde{\pi}(U)\hat{\psi}_\theta (U)\}$ by Eq.~(\ref{eq_general_lik2}) with $\tilde{\pi}(U)\hat{\psi}_\theta (U)$ included in $\tilde{v}(U)$.
Second, we have $\tilde{E}\{\tilde{\pi}(U)\hat{\psi}_\theta (U)\}=\tilde{E}\{T \hat{\psi}_\theta (U)\}$ by Eq.~(\ref{augPS_eq2}) for the augmented PS model (\ref{TS:augPS}).
Combining the two equations indicates that
$\tilde{E}[(1-T)\tilde{\pi}(U)\hat{\psi}_\theta (U)/\{1-\omega(U;\tilde{\lambda}_{1},\hat{\lambda }_2)\} ] = \tilde{E}\{T \hat{\psi}_\theta(U)\}$,
which can be easily shown to imply that $\tilde{\theta}_{\text{\scriptsize lik}}$ remains consistent when OR model (\ref{general_OR_set}) is correctly specified even if the PS model (\ref{TS:PS}) is misspecified.

The doubly robust estimators $\tilde{\theta}_{\text{\scriptsize reg}}$ and $\tilde{\theta}_{\text{\scriptsize lik}}$
can be regarded as {\it calibrated} regression and likelihood estimators,
with an important connection to calibration estimation using auxiliary information in survey sampling (\citealt{Deville1992, Tan2013}).
In fact, the estimating equations (\ref{est_reg}) and (\ref{eq_general_lik}) can be expressed as
$\sum_{1\le i\le n:\, T_i=0} w_i \,\tilde \pi(U_i) \Phi(X_i,U_i;\theta)=0$, where the (possibly negative) weights $\{w_i: T_i=0, i =1,\ldots,n\}$
are determined to satisfy the calibration equation \vspace{-.05in}
\begin{align} \label{cal_eq}
\sum_{1\le i\le n:\, T_i=0} w_i\, \tilde v(U_i) = \sum_{i=1}^n \tilde v(U_i).
\end{align}
For the likelihood estimator $\tilde{\theta}_{\text{\scriptsize lik}}$, $w_i = n^{-1} \{1-\omega(U_i;\tilde{\lambda})\}^{-1} $ by Eq.~(\ref{eq_general_lik2}).
For the regression estimator $\tilde{\theta}_{\text{\scriptsize reg}}$, it can be shown by direct calculation that
$w_i = n^{-1} \{1-\tilde \pi(U_i)\}^{-1} [1 - \tilde \kappa^\T \tilde h(U_i) / \{1-\tilde \pi(U_i)\} ]$, where
$\tilde \kappa= E^{-1}( \tilde \zeta^\T \tilde \xi) \tilde E ( \tilde\xi)  $.
However, the associated weights $w_i$ for $\tilde{\theta}_{\text{\scriptsize reg}}$ may be negative,
whereas the weights for $\tilde{\theta}_{\text{\scriptsize lik}}$ are always nonnegative by construction.
As a result, $\tilde{\theta}_{\text{\scriptsize lik}}$ tends to perform better (less likely yield outlying values) than  $\tilde{\theta}_{\text{\scriptsize reg}}$,
especially with possible PS model misspecification.
It remains an interesting but challenging topic to provide further theoretical analysis of performances of $\tilde{\theta}_{\text{\scriptsize reg}}$ and $\tilde{\theta}_{\text{\scriptsize lik}}$
in the presence of model misspecification.

\vspace{-.15in}\section{Data combination} \label{data-comb}

Consider setting (II) described in the Introduction, where another variable $Y$  in addition to $U$ is observed from the primary data.
The moment restriction model of interest is postulated in a separable form as\vspace{-.05in}
\begin{align}
\label{TS:comp_moment}
E^{(1)} \big\{ \Phi_1(Y,U;\theta)-\Phi_0(X,U;\theta)\big\}=0,
\end{align}
where $\Phi_1(y,u; \theta)$ is a vector of known functions of $(y,u,\theta)$ only whereas $\Phi_0(x,u;\theta)$ is a vector of known functions of $(x,u,\theta)$ only.
The expectation $E^{(1)} \{ \Phi_1(Y,U;\theta)\}$ can be directly estimated as simple sample averages from the primary data.
For estimation of $\theta$, the main challenge is then to estimate $E^{(1)} \{ \Phi_0(X,U;\theta) \}$ using both
the primary and secondary data, which is exactly the problem addressed in Section \ref{def}.

Similarly as in Section \ref{def}, we assume that the sample sizes $(n_1,n_0)$ are determined from binomial sampling.
The combined set of observed data are \vspace{-.05in}
\begin{align} \label{TS:data}
\{(T_i,U_i,(1-T_i)X_i, T_i Y_i): i=1, \ldots, n\},
\end{align}
where $T_i$ is an indicator variable, equal to either 1 or 0 if the $i$th unit is in the primary or auxiliary sample.
The moment conditions (\ref{TS:comp_moment}) can be represented by\vspace{-.1in}
\begin{align} \label{TS:combine_problem}
E\left\{ \Phi_1(Y,U;\theta) - \Phi_0(X,U;\theta) | T=1 \right\}=0 ,
\end{align}
Various statistical problems can be studied in the above setup of data combination as discussed by \citet{Graham2015} and references therein.

The methods and theory developed in setting (I) for moment restriction models with auxiliary data can be adopted and extended to setting (II).
An AIPW estimator $\hat{\theta}_{\text{\scriptsize NP}}$ for $\theta$ in (27) can be defined as a solution to equation similar to (9),
\begin{align} \label{general-AIPW}
\tilde E  \{ T \Phi_1(Y,U;\theta)\} - \tilde E \left[ \frac{1-T}{1-\hat\pi(U)} \hat\pi(U) \{ \Phi_0(X,U;\theta) - \hat\psi_\theta(U)\} + T \hat\psi_\theta(U) \right] =0,
\end{align}
where $\hat\pi(U)$ is a fitted propensity score using model (\ref{TS:PS}) as before,
and $\hat\psi_\theta(U)$ is a fitted outcome regression function using model (\ref{general_OR_set}), with $\Phi(x,u;\theta)$ replaced by $\Phi_0(x,u;\theta)$.
A calibrated regression estimator $\tilde{\theta}_{\text{\scriptsize reg}}$ can be defined as a solution to
\begin{align} \label{general-reg}
\tilde E  \{ T \Phi_1(Y,U;\theta)\} -  \tilde{E} \{ \tilde{\tau}_{\text{\scriptsize reg}}(\theta) \}=0,
\end{align}
and a calibrated likelihood estimator $\tilde{\theta}_{\text{\scriptsize lik}}$ defined as a solution to
\begin{align} \label{general-lik}
\tilde E  \{ T \Phi_1(Y,U;\theta)\} - \tilde{E}\left\{\frac{1-T}{1-\omega(U;\tilde{\lambda})}\tilde{\pi}(U)\Phi_0(X,U;\theta)\right\}=0 ,
\end{align}
where $\tilde{\tau}_{\text{\scriptsize reg}}(\theta) $ is defined as in (\ref{est_reg}),
and $\tilde\lambda$ defined as in (\ref{eq_general_lik2}),
with $\Phi(x,u;\theta)$ replaced by $\Phi_0(x,u;\theta)$ throughout, including the newly defined $\hat\psi_\theta(U)$ in augmented PS model (\ref{TS:augPS}).
Similarly as in Section~\ref{def}, the estimators $\hat{\theta}_{\text{\scriptsize NP}}$, $\tilde{\theta}_{\text{\scriptsize reg}}$, and $\tilde{\theta}_{\text{\scriptsize lik}}$ can be shown to be
doubly robust and locally nonparametric efficient.
Moreover,  $\tilde{\theta}_{\text{\scriptsize reg}}$  and $\tilde{\theta}_{\text{\scriptsize lik}}$
are expected to yield smaller variances than  $\hat{\theta}_{\text{\scriptsize NP}}$ and the doubly robust estimator in Graham et al.~(2016) when the propensity score model is correctly specified.
In general, $\tilde{\theta}_{\text{\scriptsize reg}}$  and $\tilde{\theta}_{\text{\scriptsize lik}}$ do not achieve intrinsic efficiency or the theoretical guarantee as in Corollary~\ref{cor-reg},
mainly due to the inefficiency of $ \tilde E  \{ T \Phi_1(Y,U;\theta)\}$ in (\ref{general-reg}) and (\ref{general-lik}),
which, however, tends to be of less concern than the variability from the second term.
It is possible to construct calibrated estimators of $\theta$ differently to achieve intrinsic efficiency, as shown in
\citet{Shu2015} for ATT estimation. But such estimators are more complex than above and may not be preferable in the case where $\Phi_1$ and $\Phi_0$ are multi-dimensional, due to finite-sample consideration.
See the end of Section~\ref{sec:reg-lik} for related discussion.

In the next section, we study two-sample instrumental variable estimation (\citealt{Klevmarken1982, Angrist1992})
as a concrete application, where estimating equations (\ref{general-AIPW})--(\ref{general-lik}) can be simplified to yield closed-form estimators.

\vspace{-.1in}
\section{Two-sample instrumental variable estimation} \label{sec:LIN_TSIV}

A typical problem in econometrics involves estimating regression coefficients in a linear regression model with endogeneity,\vspace{-.1in}
\begin{align} \label{setup}
Y =\beta X +\beta^{c \,\T} Z^c+\varepsilon,
\end{align}
where $Y$ is a response variable, $X$ is a scalar, endogenous variable possibly correlated with the error term $\varepsilon$, and $Z^c$ is a $(k-1)\times 1$ vector
of exogenous variables uncorrelated with $\varepsilon$.
Suppose that there exists a scalar instrument variable (IV) $Z$ that is correlated with $X$ but uncorrelated with $\varepsilon$.
Extension to multiple endogenous variables and instruments is possible, but would be technically more complex.
Let $\beta^\dag = (\beta, \beta^{c\,\T})^\T$ and $U = (Z, Z^{c\,\T})^\T$. Given an i.i.d. sample $\{(Y_i,X_i, U_i):i=1,\ldots,n\}$ from the primary population,
the conventional IV estimator of $\beta^\dag$  is
$
\hat \beta^\dag_{\text{\scriptsize IV}} = \tilde E^{-1}\big\{ U (X,Z^{c\,\T})\} \tilde E ( U Y)
$,
where, as before, $\tilde E()$ denotes the sample average. This estimator is identical to the two-stage least squares (2SLS) estimator, defined as
$
\hat \beta^\dag_{\text{\scriptsize 2SLS}} = \tilde E^{-1} ( \hat U \hat U^\T ) \tilde E (\hat U Y)
$,
where $\hat U = (\hat X, Z^{c\,\T})^\T$ and $\hat X = U^\T \tilde E^{-1} (U U^\T) \tilde E(U X)$.

Recently, two-sample instrumental variable estimation has been proposed to deal with the
situation where only $Y$ and $U$ (but not $X$) are observed in one primary sample,
and only $X$ and $U$ (but not $Y$) are observed in another auxiliary sample
(\citealt{Klevmarken1982, Angrist1992, Atsushi2010}).
In the notation of Section \ref{data-comb}, the two samples can be combined and represented as (\ref{TS:data}).
The moment conditions corresponding to IV estimation, \vspace{-.05in}
\begin{align} \label{eq_TSIV}
E \big\{ (Y -\beta X-\beta^{c\,\T} Z^c) U \,|\, T=1 \big\}=0,
\end{align}
can be put in the form (\ref{TS:combine_problem}), where $\Phi_1(y,u;\theta) = yu$ and $\Phi_0(x,u;\theta) = u (x,z^{c\,\T})\beta^\dag $ with $\theta=\beta^\dag$.
In the remainder of this section, we discuss existing methods in Section \ref{sec:existing-methods},
and then develop improved methods in Section \ref{sec:improved-methods}.

\vspace{-.05in}\subsection{Existing methods}  \label{sec:existing-methods}

For two-sample instrumental variable estimation, \citet{Angrist1992} used the following estimator mimicking the form of $\hat \beta^\dag_{\text{\scriptsize IV}}$:
\begin{align*} 
\hat \beta^\dag_{\text{\scriptsize TSIV}} = \Big\{ n_0^{-1} \sum_{1\le i\le n:\, T_i=0} U_i (X_i,Z_i^{c\,\T}) \Big\}^{-1} \Big( n_1^{-1} \sum_{1\le i\le n:\, T_i=1} U_i Y_i \Big) .
\end{align*}
Alternatively, another estimator motivated by two-stage least squares is defined as (e.g., \citealt{BJ1997})
\begin{align*} 
\hat \beta^\dag_{\text{\scriptsize TS2SLS}} = \Big( \sum_{1\le i\le n:\, T_i=1} \hat U_i \hat U_i^\T \Big)^{-1} \Big( \sum_{1 \le i\le n:\, T_i=1} \hat U_i Y_i \Big) .
\end{align*}
where $\hat U_i = (\hat X_i, Z_i^{c\,\T})^\T$ and $\hat X_i = U_i^\T (\sum_{1 \le j \le n: T_j=0} U_j U_j^\T )^{-1} (\sum_{1\le j \le n:T_j=0} U_j X_j)$ for $i=1,\ldots,n$.
As emphasized in \citet{Atsushi2010}, the two estimators $\hat \beta^\dag_{\text{\scriptsize TSIV}}$ and $\hat \beta^\dag_{\text{\scriptsize TS2SLS}}$ are different from each other in the two-sample case, even though
IV and 2SLS estimation are equivalent in the standard one-sample case.

The estimators $\hat \beta^\dag_{\text{\scriptsize TSIV}}$ and $\hat \beta^\dag_{\text{\scriptsize TS2SLS}}$ have been shown to be consistent provided that the two samples
are drawn from the same population, i.e., the propensity score $\pi(U)$ is a constant (\citealt{Angrist1992, Atsushi2010}).
This assumption is much more restrictive than (A2), which allows that the marginal distributions of $U$ differ
between the primary and auxiliary populations or, equivalently, the propensity score $\pi(U)$ is a nonconstant function of $U$.
Under Assumption (A2),  $\hat \beta^\dag_{\text{\scriptsize TSIV}}$ and $\hat \beta^\dag_{\text{\scriptsize TS2SLS}}$
generally become inconsistent. An interesting exception noticed by \citet{Atsushi2010} is that
$\hat \beta^\dag_{\text{\scriptsize TS2SLS}}$, but not $\hat \beta^\dag_{\text{\scriptsize TSIV}}$, remains consistent
if a linear regression model of $X$ given $U$ holds even when
the two samples are differentially stratified on $U$.
Therefore, $\hat \beta^\dag_{\text{\scriptsize TS2SLS}}$ is doubly robust (i.e., remains consistent)
if either a constant PS model is correct or a linear regression model of $X$ given $U$ is correct.
See \cite{Khanwand-Lin} and \cite{choi-etal} for more recent works, both under the assumption of
compatible moments of common variables between the two samples.

\subsection{Improved methods} \label{sec:improved-methods}

We discuss how the improved methods developed in Section \ref{def} for moment restriction models can be applied for estimating $\beta^\dag$
in the two-sample instrumental variable problem.
The application is facilitated by the fact that $\beta^\dag$ from the moment conditions (\ref{eq_TSIV}) can be represented in a closed form as
$\beta^\dag = (\mu_3, \mu_2)^{-1} \mu_1$,
where $\mu_1=E(U Y|T=1)$, $\mu_2=E(U Z^{c\,\T} | T=1)$, and $\mu_3=E( U X |T=1)$.
Then $\mu_1$ and $\mu_3$ are both $k\times1$ vectors, and $\mu_2$ is a $k\times (k-1)$ matrix, where $k$ is the dimension of $U$.
Similar ideas can be followed when the parameter of interest $\theta$ is only implicitly defined through moment conditions (\ref{TS:combine_problem}),
but will not be further pursued here.

Because $(Y,U)$ are measured in the primary data indicated by $T=1$,
the conditional means $\mu_1$ and $\mu_2$ can be directly estimated by
$\hat{\mu}_1=\tilde{E}(T UY )/\tilde{E}(T)$ and $\hat{\mu}_2=\tilde{E}(T U Z^{c\,\T})/\tilde{E}(T) $.
These estimators $\hat\mu_1$ and $\hat\mu_2$ are consistent, free of modeling assumptions.
On the other hand, $\mu_3$ can be estimated from a moment restriction model (\ref{combine_problem}),
by defining $\Phi(X,U; \theta) = U X - \mu_3$, where
$\theta=\mu_3$ is a $k \times 1$ vector of unknown parameters.
For any estimator $\hat\mu_3$, the resulting estimator of $\beta^\dag$ is
\begin{align}  \label{general_TS_est}
\hat \beta^\dag(\hat\mu_3) = (\hat\mu_3, \hat\mu_2)^{-1} \hat \mu_1.
\end{align}
The remaining task is then to estimate $\mu_3$.
Because of the use of the model-free estimators $\hat\mu_1$ and $\hat\mu_2$, consistency of $\hat \beta^\dag(\hat\mu_3)$ is directly determined by that of $\hat\mu_3$,
but efficiency properties can be subtle. See the discussion at the end of Section~\ref{sec:reg-lik}.

\subsubsection{Model-based and AIPW estimators} \label{sec:model-based}

First, simple estimators of $\mu_3$ can be derived, depending on either an outcome regression model or a propensity score model.
Suppose that a PS model (\ref{TS:PS}) is specified, and the fitted propensity score $\hat\pi(U)$ is obtained by maximum likelihood.
The IPW estimating equation (\ref{PS_est}) yields the estimator
\begin{align*}
\hat{\mu}_{3,\text{IPW}}=
\tilde{E}\left\{\frac{(1-T)\hat{\pi}(U) UX}{1-\hat{\pi}(U)}\right\}\Big/\tilde{E}\left\{\frac{(1-T)\hat{\pi}(U)}{1-\hat{\pi}(U)}\right\},
\end{align*}
which is consistent when PS model (\ref{TS:PS}) is correctly specified.
Noticing that $E(UX | U)=E(X|U)U$, consider a regression model for $E(X |U)$,
\begin{align} \label{TS:OR}
E(X|U) = m(U;\alpha)=\Psi\{\alpha^\T g(U)\},
\end{align}
where $\Psi(\cdot)$ is an inverse link function, $g(u)$ is a vector of known functions {\it including 1}, and $\alpha$ is a vector of unknown parameters.
Let $\hat\alpha$ be the least-squares estimate of $\alpha$ from the auxiliary sample (i.e., $T=0$), and let $\hat m(U) = m(U; \hat\alpha)$.
Substituting  the fitted value $\hat \psi_\theta(U)=U \hat{m}(U)-\mu_3$
 into Eq.~(\ref{OR_est}) yields the OR estimator
\begin{align*}
\hat{\mu}_{3,\text{OR}} =\tilde{E}\big\{T U \,\hat{m}(U)\big\}\big/\tilde{E}(T),
\end{align*}
which is consistent when OR model (\ref{TS:OR}) is correctly specified.
The resulting estimator $\hat\beta^\dag_{\text{\scriptsize OR}}$  from Eq.~(\ref{general_TS_est}) can be seen as a solution to the equation
\begin{align*} 
\tilde E \left[ T \left\{ Y-\beta \hat{m}(U)-\beta^{c\,\T} Z^c \right\} U \right]=0.
\end{align*}
It can be easily shown that  $\hat\beta^\dag_{\text{\scriptsize OR}}$  reduces to $\hat \beta^\dag_{\text{\scriptsize TS2SLS}} $
in the special case where model (\ref{TS:OR}) is a linear regression model of $X$ on $U$, i.e., $m(U;\alpha)=\alpha^\T U$.
In general, the two estimators $\hat\beta^\dag_{\text{\scriptsize OR}}$ and $\hat \beta^\dag_{\text{\scriptsize TS2SLS}}$ are different from each other.

Similarly, substituting $\Phi(X,U;\theta)=UX-\mu_3$ and $\hat{\psi}_\theta (U)=U \hat{m}(U) -\mu_3$ into Eq.~(\ref{est_NP}) for $\hat\theta_{\text{\scriptsize NP}}$ leads to the AIPW estimator
\begin{align*}
\hat{\mu}_{3,\text{AIPW}} =\tilde{E}\left[ \frac{1-T}{1-\hat{\pi}(U)}\hat{\pi}(U) UX-
\left\{\frac{1-T}{1-\hat{\pi}(U)}-1\right\} U \hat{m}(U) \right] \Big/\tilde{E}(T) .
\end{align*}
By Proposition \ref{prop2}, the estimator $ \hat{\mu}_{3,\text{AIPW}} $
is doubly robust, i.e., remains consistent if either PS model (\ref{TS:PS}) or OR model (\ref{TS:OR}) is correctly specified, and
$ \hat{\mu}_{3,\text{AIPW}} $ is locally nonparametric efficient, i.e., achieves
the nonparametric efficiency bound for $\mu_3$ when both PS model (\ref{TS:PS}) and OR model (\ref{TS:OR}) are correctly specified.

\subsubsection{Calibrated regression and likelihood estimators} \label{sec:reg-lik}

We apply the calibrated regression estimator $\tilde\theta_{\text{\scriptsize reg}}$ and likelihood estimator $\tilde\theta_{\text{\scriptsize lik}}$ to estimates $\mu_3$,
with $\Phi(X,U;\theta)=UX-\mu_3$ and $\hat{\psi}_\theta (U)=U \hat{m}(U) -\mu_3$.
As in Section \ref{improve_est}, assume that an augmented logistic PS model (\ref{TS:augPS}) is used,
and the fitted propensity score $\tilde \pi(X)$ is obtained by maximum likelihood.

The computation can be further simplified. First,
the additional regressors $\hat\psi_\theta (U)=U \hat{m}(U) -\mu_3$  in the augmented PS model (\ref{TS:augPS}) can be simplified as $\hat{m}(U) U$ by dropping the $\mu_3$ term,
because  $f(U)$ already includes a constant.
Similarly, the vector $\tilde{\pi}( U )\hat{\psi} (U) = \tilde\pi(U) \hat{m}(U) U - \mu_3 \tilde\pi(U)$ in $\tilde v(U)$ can be simplified
as $\tilde\pi(U) \hat{m}(U) U$ by removing the term $\mu_3 \tilde\pi(U)$.
Then $\tilde{h}(U) = \{\tilde h_1^\T(U), \tilde h_2^\T(U)\}$ is redefined with
$\tilde{h }_1( U )= \tilde{\pi}(U)\tilde{v } ( U )$, $\tilde{v}(U)= \{ \tilde{\pi}( U ),\tilde{\pi}( U )\hat{m}(U )U^\T \}^\T$,
and $\tilde{h }_2(Z )=\tilde{\pi}(U) \{1-\tilde{\pi}(U)\} \{f^\T(U), $ $\hat{m}(U) U^\T\}^\T$.
Therefore, $\tilde\pi(U)$ and $\tilde h(U)$ are independent of $\theta=\mu_3$.
As mentioned in Section~\ref{sec:cal-reg}, $\tilde h_2$ can also be removed from $\tilde h$ for finite-sample considerations.

Substituting these definitions into Eq.~(\ref{est_reg}) yields the regression estimator
\begin{align} \label{TSIV_reg}
\tilde{\mu}_{3,\text{reg}}=\frac{\tilde{E}(\tilde{\eta}-\tilde{\beta}^{\T}\tilde{\xi})}{\tilde{E}(\tilde{\rho}-
\tilde{\kappa}^{\T}\tilde{\xi})},
\end{align}
where $\tilde{\eta}=[(1-T)/\{1-\tilde{\pi}(U)\}]\tilde{\pi}(U)UX$,
$\tilde{\beta}= \tilde{E}^{-1} (\tilde{\xi}\tilde{\zeta}^{\T}) \tilde{E} (\tilde{\xi}\tilde{\eta}^{\T})$,
$\tilde{\rho}=[(1-T)/\{1-\tilde{\pi}(U)\}]\tilde{\pi}(U)$,
and $\tilde{\kappa}= \tilde{E}^{-1} (\tilde{\xi}\tilde{\zeta}^{\T}) \tilde{E} (\tilde{\xi}\tilde{\rho}^{\T})$.
But Eq.~(\ref{TSIV_reg}) can be further simplified to
\begin{align*}
\tilde{\mu}_{3,\text{reg}}=\tilde{E} (\tilde{\eta}-\tilde{\beta}^{\T}\tilde{\xi} )\big/\tilde{E}(T),
\end{align*}
because $\tilde{E}(\tilde{\rho}-\tilde{\kappa}^{\T}\tilde{\xi}) = \tilde E \{ \tilde\pi(U)\} = \tilde E(T)$.
The first equality can be shown by direct calculation using the fact that $\tilde\rho$ is a component of $\tilde\zeta$ because $\tilde \pi(U)$ is included as a component of $\tilde v(U) = \tilde h_1(U)/\tilde \pi(U)$.
The second equality holds by the score equation (\ref{ps_score}) because a constant is included in $f(U)$.

Substituting $\Phi(X,U;\mu_3)=UX-\mu_3$ into Eq.~(\ref{eq_general_lik}) yields the likelihood estimator
\begin{align}
\tilde{\mu}_{3,\text{lik}}&=\tilde{E}\left\{\frac{1-T}{1-\omega(U ; \tilde{\lambda})}\tilde{\pi}(U) U X \right\} \Big/\tilde{E}\left\{\frac{1-T}{1-\omega(U ; \tilde{\lambda})}\tilde{\pi}(U)\right\} \nonumber \\
&=\tilde{E}\left\{\frac{1-T}{1-\omega(z ; \tilde{\lambda})}\tilde{\pi}(U) UX \right\}\Big/\tilde{E}(T),
\end{align}
where the second equality holds due to Eq.~(\ref{eq_general_lik2}) with $\tilde{\pi}(U)$ included in $\tilde{v}(U)$ and to $\tilde{E}\{ T-\tilde{\pi}(U)\}=0$ by  the score equation (\ref{augPS_eq}).

By Propositions \ref{prop-reg} and \ref{prop-lik}, the estimators $\tilde{\mu}_{3,\text{reg}}$ and $\tilde{\mu}_{3,\text{lik}}$
are not only doubly robust and locally nonparametric efficient similarly as $\hat {\mu}_{3,\text{AIPW}}$,
but also intrinsically efficient within a class of estimators as solutions to Eq.~(\ref{theta_class}), including
the estimators $\hat {\mu}_{3,\text{IPW}}$ and $\hat {\mu}_{3,\text{AIPW}}$
but with $\hat\pi(U)$ replaced by $\tilde\pi(U)$.

Finally, we examine statistical properties for the resulting estimators of $\beta^\dag$, denoted by
$\hat\beta^\dag_{\text{\scriptsize AIPW}}=\hat\beta^\dag(\hat {\mu}_{3,\text{AIPW}})$, $\tilde \beta^\dag_{\text{\scriptsize reg}}=\hat\beta^\dag(\tilde {\mu}_{3,\text{reg}})$,
or $\tilde \beta^\dag_{\text{\scriptsize lik}}=\hat\beta^\dag(\tilde {\mu}_{3,\text{lik}})$ by Eq.~(\ref{general_TS_est}).
The estimators $\hat\mu_1$ and $\hat\mu_2$ are only based on sample averages and hence are fully robust (consistent without modeling assumptions) and nonparametrically efficient
(achieving the nonparametric efficiency bounds), as shown in \citet{Shu2015} in the context of ATT estimation.
As a result, the estimators $\hat\beta^\dag_{\text{\scriptsize AIPW}}$, $\tilde \beta^\dag_{\text{\scriptsize reg}}$, and $\tilde \beta^\dag_{\text{\scriptsize lik}}$
are doubly robust, i.e., remain consistent if either PS model (\ref{TS:PS}) or OR model (\ref{TS:OR}) is correctly specified, and are locally nonparametric efficient, i.e., achieve
the nonparametric efficiency bound for $\mu_3$ when both PS model (\ref{TS:PS}) and OR model (\ref{TS:OR}) are correctly specified.
On the other hand, $\tilde \beta^\dag_{\text{\scriptsize reg}}$ and $\tilde \beta^\dag_{\text{\scriptsize lik}}$ do not generally inherit the intrinsic efficiency of
$\tilde{\mu}_{3,\text{reg}}$ and $\tilde{\mu}_{3,\text{lik}}$, because
$\hat\mu_1$ and $\hat \mu_2$ are not intrinsically efficient.
Although it is possible to derive regression and likelihood estimators of $\mu_1$ and $\mu_2$ to achieve double robustness and intrinsic efficiency,
such estimators are no longer fully robust and, more importantly, may not perform well in small or moderate samples
because they would depend on augmentation of PS model (\ref{TS:PS}) with additional regressors
as many as the dimensions, $k\times 1$ and $k \times (k-1)$  of $UY$ and $U Z^c$.
Nevertheless, by the intrinsic efficiency of $\tilde{\mu}_{3,\text{reg}}$ and $\tilde{\mu}_{3,\text{lik}}$,
the estimators  $\tilde \beta^\dag_{\text{\scriptsize reg}}$ and $\tilde \beta^\dag_{\text{\scriptsize lik}}$ are still often
more efficient than $\hat\beta^\dag_{\text{\scriptsize AIPW}}$ when PS model (\ref{TS:PS}) is correctly specified but
OR model (\ref{TS:OR}) is misspecified, as demonstrated in our simulation studies.

\section{Simulation study} \label{sec:simulation}

We conducted a simulation study to compare various estimators for two-sample IV estimation as discussed in Section \ref{sec:LIN_TSIV}.
For the primary population, suppose that the response variable $Y$ is defined as
$$
Y = 0.5 X -0.4 Z_1 + 0.5 Z_2+ \varepsilon,
$$
the endogenous variable $X$ is defined as
\begin{align} \label{x_primary}
X = Z_0 + 0.6 Z_1 -0.5 Z_2+ e,
\end{align}
where $(Z_0,Z_1,Z_2)$ are mutually independent and marginally distributed as $\N(1,1)$, and $(\varepsilon,e)$ are distributed
independently of $(Z_0,Z_1,Z_2)$ as
\begin{equation*}
\begin{pmatrix}
\varepsilon\\
e
\end{pmatrix}
\sim \N\left\{
\begin{pmatrix}
0\\
0
\end{pmatrix},
\begin{pmatrix}
1   &0.8\\
0.8 &1
\end{pmatrix}
\right\} .
\end{equation*}
In the notation of Section \ref{sec:LIN_TSIV}, the instrumental variable is $Z = Z_0$, the vector of exogenous variables is $Z^c =(Z_1,Z_2)^\T$,
and hence $U = (Z_0, Z_1, Z_2)^\T$.
For the auxiliary population, suppose that $X$ is also defined by Eq.~(\ref{x_primary}),
but $(Z_0,Z_1,Z_2,e)$ are mutually independent and marginally distributed as $\N(0,1)$.

For each simulation, we generated an i.i.d.~sample of size $n_1=5000$ and recorded only $(Y,U)$ from the primary population,
and generated an i.i.d.~sample of size $n_0=500$ and recorded only $(X,U)$ from the auxiliary population.
The two samples are then merged into one, and an indicator variable $T$ is introduced, equal to 1 or 0 for the primary or auxiliary sample respectively.
By direct calculation based on the density ratio of $U$ between the two samples, the true propensity score is
\begin{align*} 
P(T=1|U)=\text{expit} ( -1.5 +\log10+Z_0+Z_1+Z_2 ).
\end{align*}
See the Supplmentary Material for simulation results where the coefficient of the instrument $Z_0$ in Eq.~(\ref{x_primary}) is smaller, $.8$ or $.6$,
corresponding to a weaker instrument, or where the primary and auxiliary data sizes are reversed: $n_1=500$ and $n_0=5000$.
The relative performances of the estimates of $\beta$ are similar as discussed below.

\begin{figure}
\caption{Scatterplots of $X$ versus transformed variables in the auxiliary data and boxplots of transformed variables between the two samples ($n_1=5000$, $n_0=500$)}
\label{scatterplot}
\begin{tabular}{c}
\includegraphics[width=4in, height=5.7in, angle=270]{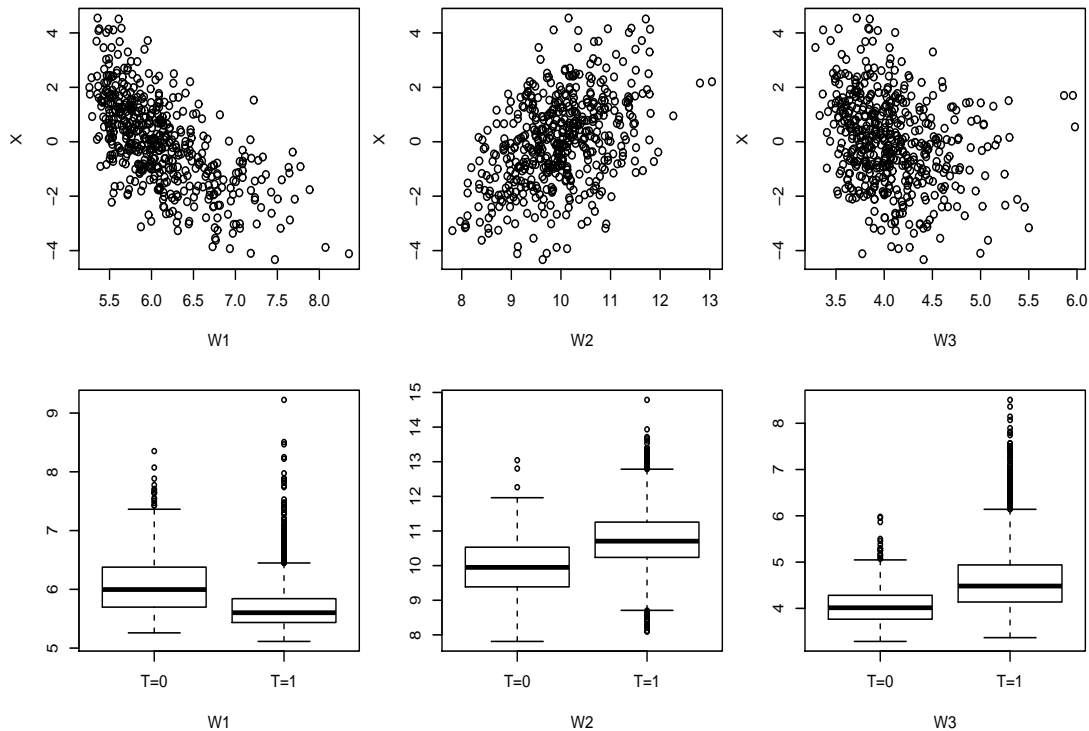}
\end{tabular}
\end{figure}

To investigate possible model misspecification similarly as in \cite{KS2007}, we define the transformed variables
\begin{align*}
& W_0=\exp(-0.5 Z_0)+5,\quad
W_1=Z_1/\{1+0.1\exp(Z_0)\}+10, \quad
W_2=\exp(0.4 Z_2)+3.
\end{align*}
We constructed the OR model (\ref{TS:OR}) with the identity link $\Psi(\cdot)$ and $g(U)$ set to either $(1,Z_0, Z_1, Z_2)$ or $(1, W_0, W_1, W_2)$,
corresponding to a correct and misspecified OR model respectively. Similarly, we constructed the PS model (\ref{TS:PS}) with the logistic link $\Pi(\cdot)$ and
$f(U)$ set to $(1,Z_0, Z_1, Z_2)$ or $(1, W_0, W_1, W_2)$, corresponding to a correct and misspecified PS model respectively.
Figure \ref{scatterplot} shows the scatterplots of $X$ versus the transformed variables $(W_0,W_1,W_2)$ in the auxiliary data and
the boxplots of $(W_0,W_1,W_2)$ between the two samples.
From these plots, the OR model and PS models based on the transformed variables seem to be reasonable from the usual viewpoint of data analysis, even though they are misspecified.

\begin{figure}
\caption{\small Boxplots of estimates of $\beta$ relative to $0.5$ ($n_1=5000$, $n_0=500$)} \vspace{.1in}
\label{plot:TSIVboxplots}
\begin{tabular}{c}
\includegraphics[width=5.7in, height=4in]{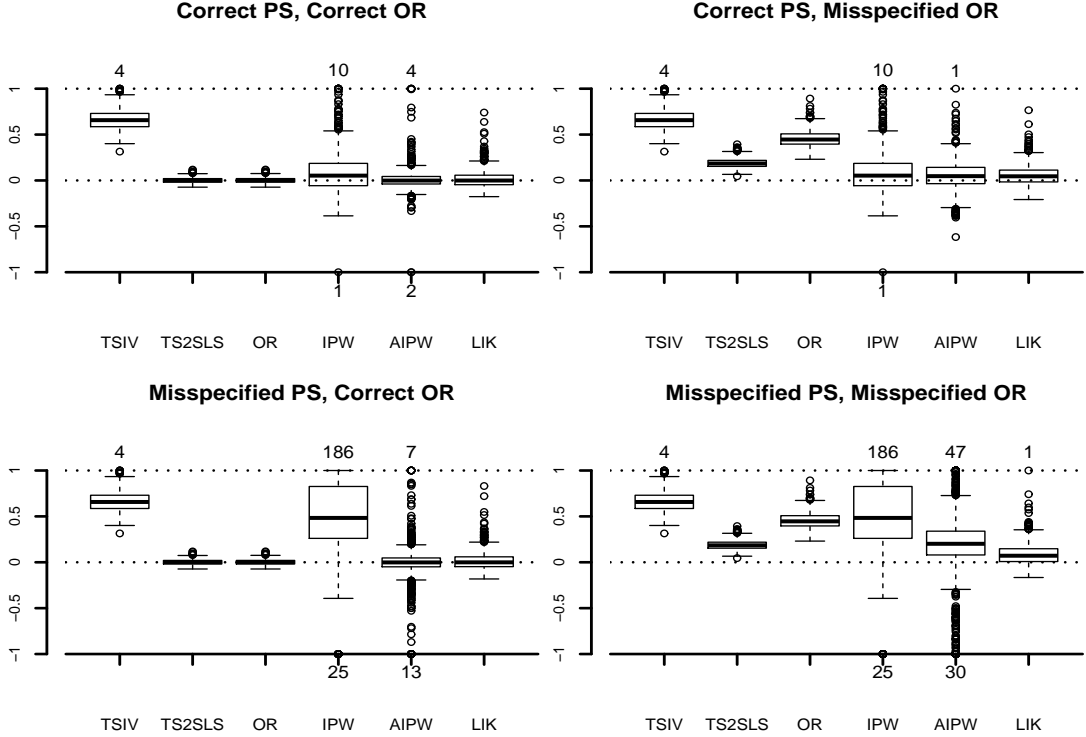}
\end{tabular} \vspace{-.1in}
\end{figure}

Table \ref{simu:comp} summarizes the results based on 1000 repeated simulations for various estimates of $\beta$ for two-sample IV estimation under four scenarios of OR and PS model specification.
Figure \ref{plot:TSIVboxplots} shows the boxplots of the differences between the estimates of $\beta$ and the true value $0.5$. The realizations of each estimator are censored within the range of the $y$-axis, and the number of realizations that lie outside the range are marked next to the lower or the upper limit of $y$-axis for each estimator.

 \begin{table}
   \caption{Summary of estimates of $\beta$ relative to the truth $0.5$ ($n_1=5000$, $n_0=500$)}
   \scriptsize
   \label{simu:comp}
   \begin{center}
   \begin{tabular}{lcccccc}
   \hline\hline
\noalign{\medskip}
                       & TSIV      & TS2SLS    & OR        & IPW & AIPW      & LIK       \\
\noalign{\medskip}\hline\noalign{\medskip}
Correct PS, Correct OR & 0.66330	& 0.00119	& 0.00138	& 0.09151	& 0.01079	& 0.01514 \\
                       & (0.10858)	& (0.02886)	& (0.02891)	& (0.42858)	& (0.15080)	& (0.09404) \\
                       \noalign{\medskip}
Correct PS, Misspecified OR   & 0.66330	& 0.18837	& 0.45569	& 0.09151	& 0.05283	& 0.05582  \\
                       & (0.10858)	& (0.05056)	& (0.08390)	& (0.42858) & (0.16045)	& (0.10712) \\
                       \noalign{\medskip}
Misspecified PS, Correct OR   & 0.66330	& 0.00119	& 0.00138	& 0.63761	& -0.01244	& 0.01656  \\
                       & (0.10858)	& (0.02886)	& (0.02891)	& (2.19874)	& (0.42315)	& (0.09916) \\
                       \noalign{\medskip}
Misspecified PS, Misspecified OR    & 0.66330	& 0.18837	& 0.45569	& 0.63761	& 0.04598	& 0.08604  \\
                       & (0.10858)	& (0.05056)	& (0.08390)	& (2.19874)	& (3.26092) & (0.11847) \\
\noalign{\medskip}\hline\hline
\end{tabular}
   \end{center}
   \vspace{-0.1in}
The estimators of $\beta$ are taken from $\hat\beta^\dag_{\text{\tiny TSIV}}$, $\hat\beta^\dag_{\text{\tiny TS2SLS}}$, and $\hat\beta^\dag$ in Eq.~(\ref{general_TS_est})
where $\hat\mu_3$ is set to $\hat\mu_{3,\text{OR}}$, $\hat\mu_{3,\text{IPW}}$, $\hat\mu_{3,\text{AIPW}}$, and $\tilde\mu_{3,\text{lik}}$
with $\tilde h_2$ removed from $\tilde h$ (see Sections~\ref{sec:model-based}-\ref{sec:reg-lik}).
The results for the estimator of $\beta$ based on $\tilde\mu_{3,\text{reg}}$ are less satisfactory than based on $\tilde\mu_{3,\text{lik}}$ and hence not shown.
Each cell gives the empirical bias (upper) and standard deviation (lower) of the point estimator, from 1000 Monte Carlo samples.
   \end{table}

\normalsize

The following remarks be drawn on the comparisons of various estimators.\vspace{-.1in}
\begin{itemize}\addtolength{\itemsep}{-.1in}
\item The TSIV estimator (\citealt{Angrist1992}) does not depend on the PS model or OR model used, but it yields dramatic biases in all the four scenarios,
because the common variables $U$ are distributed differently between the two samples.
As discussed in Section \ref{sec:existing-methods}, the TSIV estimator is generally consistent only when the two samples are drawn from the same population.

\item The TS2SLS and OR estimators are equivalent because $g(U) = (1,Z_0,Z_1,Z_2)$ as discussed in Section \ref{sec:model-based}, and both are approximately unbiased, when OR model is correctly specified.
On the other hand, these two estimators differ from each other, and both become biased, when OR model is misspecified.

\item The IPW estimator is approximately unbiased only when PS model is correctly specified and it has very large variances whether PS model is correctly specified or misspecified.
Such a performance is typical of simple IPW estimators.

\item The AIPW estimator and LIK estimator are doubly robust: they are approximately unbiased when either PS model or OR model is correctly specified.
As implied by local efficiency, the two estimators have similar variances to each other when both PS model and OR model are correctly specified.
However, when PS model is correctly specified but OR model is misspecified, the LIK estimator has a smaller variance than AIPW by a factor about $(.160/.107)^2=2.24$,
due to intrinsic efficiency of the calibrated likelihood estimator used for estimating $\mu_3= (UX|T=1)$, the central quantity in two-sample IV estimation.

\item The LIK estimator also has a much smaller variance than AIPW, which yields a considerable number of outlying values, when the OR model is correctly specified but PS model is misspecified,
although the two estimators are approximately unbiased in this scenario.
This comparison shows that the LIK estimator can be less sensitive than AIPW to misspecification of the PS model.
\end{itemize}

\section{Reassessing public housing projects}

In order to improve the quality of housing and prospects of children in poor families, the US federal government has provided substantial housing assistance such as public housing projects to low-income families since 1937. However, public dissatisfaction with public housing projects remained high, largely in response to the rising cost of public housing and the high rates of crime, unemployment and school failure among public housing residents. But there was little evidence, beyond newspaper accounts, on the negative impact of public housing on children. In this context, \citet{Currie2000} investigated the effects of participation in public housing projects on the living quality and children's educational attainment, using two-sample IV analysis which combines information from different data sources.

\citet{Currie2000} restricted the analysis to families with exactly two children under 18 in the household, for reasons as discussed later related to the validity of the instrumental variable used.
To study the effects of project participation on various outcomes, the linear regression model (\ref{setup}) can be expressed as
\begin{align} \label{setup2}
\mbox{OUTCOME} = \beta * \mbox{PROJ} + \beta^{c\,\T} *\mbox{EXOG} + \varepsilon,
\end{align}
where PROJ is project participation ($X$), defined as 1 if a family lived in a housing project or 0 otherwise, and EXOG includes exogenous explanatory variables ($Z^c$) such as the household head's gender, age, race, education, marital status and the number of boys in the family and so on.
The OUTCOME variable can be a measure of housing quality (overcrowding or housing density) or children's educational attainment (grade repetition).
For simplicity, we take ``overcrowdedness" as the outcome of interest ($Y$), which is defined as $1$ if a family had two or less living/bedrooms
and $0$ otherwise.
See the Supplementary Materials for the results with housing density as the outcome.
An important aspect of model~(\ref{setup2}) is that PROJ is considered an endogenous variable, possibly correlated with the error term $\varepsilon$,
due to unobserved factors affecting both project participation and outcomes. In fact, families are eligible in projects if they had incomes at or below 50\% of the area median. They may be more likely to live in substandard housing and their children may be more likely to experience negative outcomes, even if not participating in projects.

To control for the endogeneity, \citet{Currie2000} identified sex composition as an instrument ($Z$) for project participation, defined as 1 if a family had a boy and a girl
and 0 if two boys or two girls.
Under the Department of Housing and Urban Development (HUD) rules, boys and girls cannot be required to share one bedroom except very young children. As a result, a family with two boys or two girls would be eligible for a two-bedroom apartment, whereas a family with a boy and a girl would be eligible for a three-bedroom apartment.
In order for sex composition to be a valid instrument, it should influence project participation $X$, but
have no direct effect on the outcome, overcrowding, except through $X$.
To abstract from any effects due to the number of children, \citet{Currie2000} restricted the analysis to families with exactly two children under 18.
In this case, families with a boy and a girl should be more likely to participate in projects, by the benefit of an extra room.
On the other hand, \citet{Currie2000} pointed out that there was little reason to expect sex composition would affect overcrowding at least as they defined.
In fact, families with two children of opposite sex would probably seek a change from two to three bedrooms,
but not a change from one ($Y=1$) to two ($Y=0$) bedrooms.

\begin{figure}[t]
\begin{center}
\caption{\small Sample averages of the common variables $U$ between the two samples} \vspace{-.2in}
\label{boxplot:var}
\begin{tabular}{c}
\includegraphics[width=4in, height=5.7in, angle=270]{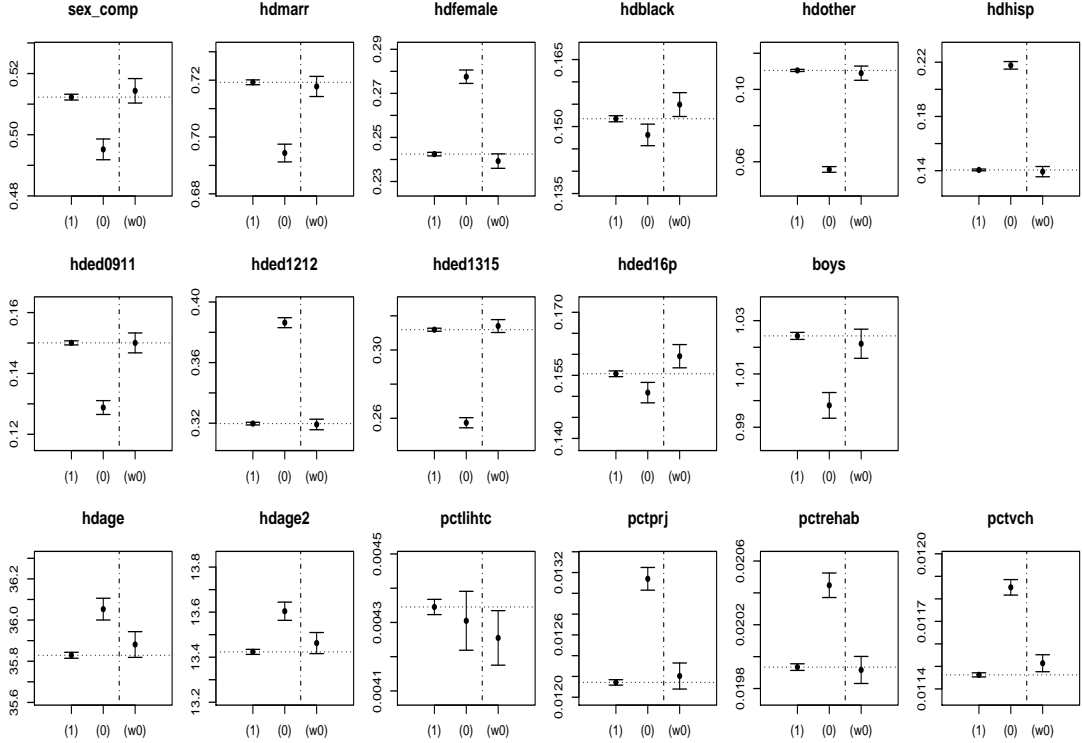}
\end{tabular}
\end{center} \vspace{.1in}
\scriptsize
Label ``(1)" or ``(0)" denotes, respectively, the simple sample means from the Census or CPS data, and ``(w0)" denotes the inverse probability weighed means from the CPS data.
\end{figure}

\citet{Currie2000} showed that living in projects is associated with poorer outcomes, using data from the Survey of Income and Program Participation (SIPP).
But they found that the SIPP sample is too small to yield reliable estimates using standard IV methods. Therefore, \citet{Currie2000}
used the two-sample IV method to combine information from 1990 Census data and 1990 to 1995 waves of the March Current Population Survey (CPS).
The Census data of size $n_1=279129$ are the \textit{primary} data, which contain the outcome $Y$ for overcrowding, the instrumental variable $Z$ for sex composition,
and exogenous explanatory variables $Z^c$ as seen from Table 4 of \citet{Currie2000}.
The CPS data of size $n_0=21718$ are the \textit{auxiliary} data, which contain the dummy variable $X$ for housing project participation and $(Z, Z^c)$ defined exactly the same as in Census data.

Figure \ref{boxplot:var} shows the simple sample means with error bars (one standard error) for all the common variables $(Z, Z^c)$ from the Census and CPS samples.
The binary variables in the first two rows include information about the household head's marital status, gender, race, education (``hdmarr",``hdfemale",``hdblack", etc.).
The continuous variable in the third row include the age of household head (``hdage") and its squared value ``hdage2", the percentage of households in projects or other subsidized housing (``pctprj"), and so on.
Except ``hdblack", ``hded16p" and ``pctlihtc", all the variables have significantly different means at 5\% level between the two samples. Therefore, the two samples are representative of different populations.
In this situation, the TSIV estimator in \citet{Angrist1992} is generally biased, as discussed in Section \ref{sec:existing-methods} and illustrated in the simulation study.

\begin{figure}[t]
\begin{center}
\caption{\small Histograms of fitted propensity scores from model (\ref{TS:PS})} \vspace{-.2in}
\label{ppi}
\begin{tabular}{c}
\includegraphics[width=2in, height=4in, angle=270]{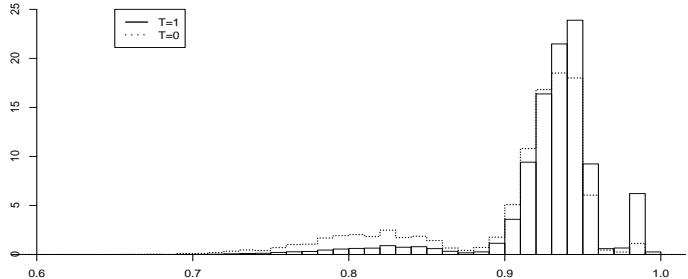}
\end{tabular}
\vspace{-0.2in}
\end{center}
\end{figure}

We apply the six estimators as studied in Section \ref{sec:simulation} to estimate $\beta$ in model (\ref{setup2}).
The OR model (\ref{TS:OR}) is specified with the  identity link and regressors $g(U)=(1,Z,Z^{c\,\T})^\T$.
The PS model (\ref{TS:PS}) is specified with the logistic link and $f(U)$ including 1, the main effects $(Z, Z^{c})$, and the interaction $\texttt{hdother:hdhisp}$.
Initially, the PS model with only the main effects is fitted, and
covariate balance is examined by comparing the sample means from Census data and inverse probability weighted means from CPS data (\citealt{Rosenbaum1984}).
The interaction term is then added to improve covariate balance as shown in Figure~\ref{boxplot:var}.
Figure \ref{ppi} presents the histograms of the fitted propensity scores $\hat\pi(U)$ separately for the two samples.
As is consistent with Figure \ref{boxplot:var}, the fitted propensity scores vary noticeably, about $n_1/(n_1+n_0)=0.928$, indicating
that the two samples are likely to be drawn from different populations.

\begin{figure} \label{boxplots:housing}
\caption{\small Boxplots of bootstrap estimates of $\beta$ (overcrowdedness)} \vspace{-.1in}
\begin{tabular}{c}
\includegraphics[width=5.7in, height=2in]{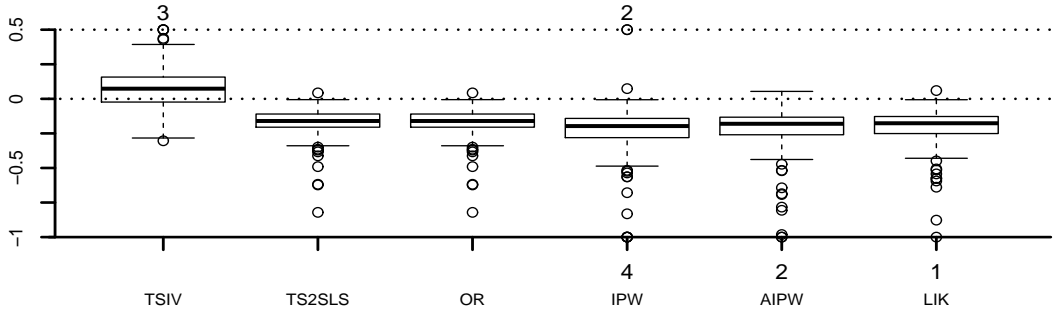}
\end{tabular}
\vspace{-0.1in}
\end{figure}

Table \ref{results of real data} presents the point estimates of $\beta$, the analytical and bootstrap standard errors, and
95\% bootstrap confidence intervals. Bootstrap is performed by drawing bootstrap samples separately from the Census and CPS data in our two-sample setting.
Figure \ref{boxplots:housing} shows
the boxplots of the six estimators of $\beta$ from 200 bootstrap samples.
Similarly as in Figure \ref{plot:TSIVboxplots}, the number of realizations that lie outside the range of $y$-axis are marked next to the lower or the upper limit of $y$-axis for each estimator.

We obtain the same TS2SLS point estimate as \citet{Currie2000}, $-0.1595$, showing the households in public housing projects are less likely to be overcrowded.
Because $g(U)=(1,Z,Z^{c\,\T})^\T$, the OR estimate is identical to TS2SLS as discussed in Section \ref{sec:model-based}.
Our bootstrap standard error is somewhat larger than the analytical standard error reported in \citet{Currie2000}.
The 95\% bootstrap percentile confidence interval still falls to the left of 0, confirming
that the effect of housing projects is likely to be positive in reducing overcrowdedness.

The TSIV estimate is positive with 95\% confidence interval covering 0. But this result is probably biased as discussed above.
Compared with the TS2SLS results,
the IPW and AIPW estimates are associated with much larger standard errors, more serious outlying values, and wider confidence intervals.
The LIK point estimate is between the TS2SLS and IPW/AIPW estimates,
with bootstrap standard error also between the corresponding standard errors.
The 95\% bootstrap confidence interval from LIK, however, still falls to the left of 0, indicating that
participation in housing projects could alleviate the overcrowdedness of households.
This result agrees with that from TS2SLS, but can be seen to be more robust, allowing that either the linear regression in the first stage of TS2SLS is valid or
the propensity score model is correctly specified for capturing differences between the two samples.

\begin{table}
   \caption{Estimates of the effect of project participation on overcrowdedness} \vspace{-.1in}
   \scriptsize
   \label{results of real data}
   \begin{center}
   \begin{tabular}{lcccccc}
  \hline\hline
\noalign{\medskip}   &TSIV     &TS2SLS      &OR       &IPW     &AIPW            &LIK\\
\noalign{\smallskip} \hline   \noalign{\smallskip}
 Point       &0.06900	& -0.15949	& -0.15949	& -0.20437	& -0.18821	& -0.18289\\
 SE       &0.12834	& ---	& 0.06241	& 0.11430	& 0.09785	& 0.09494 \\
 boot.SE       &0.15408	& 0.10138	& 0.10138	& 3.57790	& 5.17650	& 0.18207\\
 boot.CI       &(-0.220,0.394)	& (-0.384,-0.049)	& (-0.384,-0.049)	& (-0.683,-0.033)	& (-0.694,-0.043)	& (-0.575,-0.043) \\ \noalign{\smallskip}
         \hline\hline
\end{tabular}
   \end{center}
\vspace{-0.1in}
Each column gives the point estimate (upper), the analytical and bootstrap (boot) standard errors (middle), and 95\% bootstrap percentile confidence interval (lower) from 200 bootstrap samples.
For comparison, the TS2SLS estimate is reported as $-0.1595$, with analytical standard error $0.0624$, in \citet{Currie2000}.
   \end{table}

\vspace{-.1in}
\section{Conclusion}

Combining information from two or more samples is important to various biomedical and economic studies.
However, these samples may often differ in the marginal distributions of the common variables.
Such differences need to be appropriately accounted for in order to draw valid and accurate inferences from the combined data.

We develop various estimators with improved statistical properties
for moment restriction models with data combination from two samples, provided that the distributions of the missing data given the common variables are the same between the two samples.
As a concrete application, we study the two-sample instrumental variable problem. The simulation study and reanalysis of public housing projects demonstrate the advantage of our estimators compared with existing estimators.

\setlength{\bibsep}{1pt}

\bibliographystyle{myapa}
    \bibliography{thesisnew2}


\clearpage
\setcounter{page}{1}
\setcounter{section}{0}

\setcounter{equation}{0}

\setcounter{pro}{0}
\renewcommand{\thepro}{S\arabic{pro}}

\renewcommand{\theequation}{S\arabic{equation}}
\renewcommand{\thesection}{\Roman{section}}
\renewcommand\thefigure{S\arabic{figure}}
\renewcommand\thetable{S\arabic{table}}

\setcounter{figure}{0}
\setcounter{table}{0}

\centerline{\bf\large Supplementary Material}
\centerline{\bf for ``Improved Methods for Moment Restriction Models with Marginally}
\centerline{\bf Incompatible Data Combination and an Application to Two-sample}
\centerline{\bf Instrumental Variable Estimation" by Shu \& Tan}
\vspace{.1in}

\section{Review of semiparametric theory}

Proposition \ref{prop1} restates the efficient influence functions from \citeappend{Chen2008} for estimation of $\theta_0$ under moment conditions (\ref{combine_problem}) in three different settings.

\footnotetext[1]{It is originally assumed in \citeappend{Chen2008} that $0<P(T=1|U=u)<1$ for all $u$. However, the proofs in \citeappend{Chen2008} can be seen to still hold even when $P(T=1|U=u)$ is 0 for some values $u$, because only subjects with $T=0$ are inversely weighted by $1-\pi(U)$.}

\begin{pro}[\citealtappend{Chen2008}] \label{prop1}
Let $q = P(T=1)$,
$\Gamma_{\theta}=(\partial/\partial\theta^{\T})E \{ \Phi(X,U;\theta)| $ $T =1 \}$, and
\begin{align*}
F(T,X,U; \theta) &=\frac{1-T}{q}\frac{\pi(U)}{1-\pi(U)} \{\Phi(X,U;\theta)-\psi_\theta (U)\},
\end{align*}
where $\pi(U) = P(T=1|U)$ and $\psi_\theta(U)=E\{ \Phi(X,U;\theta)| U\}$.
Under Assumptions (A1)--(A3),\footnotemark[1] the efficient influence function for estimation of $\theta_0$ is as follows.
\begin{enumerate}
\item[(i)]The efficient influence function is
$$
\varphi_{\text{\scriptsize NP}}(T,X,U)=-\Gamma_{\theta_0}^{-1}\times F_{\text{\scriptsize NP}}(T,X,U;\theta_0),
$$
where $F_{\text{\scriptsize NP}}(T,X,U;\theta)=F(T,X,U; \theta)+ T \psi_\theta(U) /q$.

\item[(ii)]If the propensity score $\pi(U)$ is known, then the efficient influence function is
\begin{align*}
\varphi_{\text{\scriptsize SP*}}(T,X,U)=-\Gamma_{\theta_0}^{-1}\times F_{\text{\scriptsize SP*}}(T,X,U;\theta_0),
\end{align*}
where $F_{\text{\scriptsize SP*}}(T,X,U;\theta)=F(T,X,U; \theta)+ \pi(U)\psi_\theta(U) /q$.

\item[(iii)]If the propensity score $\pi( U )$ is unknown but assumed to belong to a parametric family $\pi(U;\gamma)$, then the efficient influence function is
$$
\varphi_{\text{\scriptsize SP}}(T,X,U)=-\Gamma_{\theta_0}^{-1}\times F_{\text{\scriptsize SP}}(T,X,U;\theta_0),
$$
where
$F_{\text{\scriptsize SP}}(T,X,U;\theta)=F_{\text{\scriptsize SP*}}(T,X,U;\theta)+\mbox{Proj} \big[\big\{T-\pi(U)\big\}\frac{\psi_\theta(U)}{q}\big|S_{\gamma}(T,U) \big]$.
\end{enumerate}
For two random vectors $U_1$ and $U_2$, $\mbox{Proj}(U_2 | U_1) = \cov(U_2,U_1) \var^{-1}(U_1) U_1$, i.e., the linear projection of $U_2$ onto $U_1$.
\end{pro}

As discussed in \citetappend{Chen2008}, the semiparametric efficiency bounds satisfy
$V_{\text{\scriptsize NP}} \ge V_{\text{\scriptsize SP}} \ge V_{\text{\scriptsize SP*}} $,
with strict inequalities holding in general, where
$V_{\text{\scriptsize NP}}$, $V_{\text{\scriptsize SP}}$, and
$V_{\text{\scriptsize SP*}}$ are respectively the variances of $\varphi_{\text{\scriptsize NP}}$, $\varphi_{\text{\scriptsize SP}}$, and $\varphi_{\text{\scriptsize SP*}}$.
This ordering of efficiency bounds agrees with the usual comparison that the efficiency bound under a more restrictive model is no greater than under a less restrictive model.
But this relationship between the efficiency bounds, $V_{\text{\scriptsize NP}}, V_{\text{\scriptsize SP}}, V_{\text{\scriptsize SP*}}$,
differs from the situation where the efficiency bounds remain the same when the propensity score is unknown, or assumed in a parametric family, or known,
in various other missing-data problems (e.g., \citealtappend{Robins1994, Tsiatis2006}), such as the ``verify-in-sample'' case in \citeappend{Chen2008}.
This difference is related to the fact that the propensity score is ancillary for estimation of ATE, but not ancillary for estimation of ATT (\citealtappend{Hahn1998}).

\section{Technical details} \label{tech_details}

\subsection{Preparation}

Throughout, we make the following assumptions regarding the estimators $\hat \alpha$ for OR model (\ref{general_OR_set}), $\hat\gamma$ for PS model (\ref{TS:PS}),
and $(\tilde\gamma,\tilde\delta)$ for augmented PS model (\ref{TS:augPS}), allowing for possible model misspecification (e.g., \citealtappend{White1982}).

\begin{enumerate}
\item[(C1)] Assume that
$\hat\alpha$ converges to a constant $\alpha^*$ such that $\hat\alpha - \alpha^* = O_p(n^{-1/2})$.
Write $\psi^*_\theta(U) = \psi_\theta(U; \alpha^*)$.
If model (\ref{general_OR_set}) is correctly specified, then $\psi^*_\theta(U) = \psi_\theta(U)$. In general, $\psi^*_\theta(U)$ and $\psi_\theta(U)$ may differ from each other.

\item[(C2)] Assume that $\hat\gamma$ converges to a constant $\gamma^*$ such that
\begin{align*}
\hat \gamma - \gamma^* = V^{-1} \, \tilde E \left\{ s_{\gamma^*}(T,U) \right\} + o_p(n^{-1/2}),
\end{align*}
where $E\{s_{\gamma^*}(T,U)\}=0$, and the matrix $V= - E\{ \partial s_\gamma(T,U)/\partial \gamma^\T\} | _{\gamma=\gamma^*}$ is nonsingular.
Write $\pi^*(U) = \pi(U; \gamma^*)$.
If model (\ref{TS:PS}) is correctly specified, then $\pi^*(U) = \pi(U)$
and $V = \var \{ s_{\gamma^*}(T,U)  \}$. In general, $\pi^*(U)$ and
$\pi(U)$ may differ from each other.

\item[(C3)] For augmented PS model (\ref{TS:augPS}), define
$$
s^\dag(T,U; \gamma,\delta, \alpha)= \{T- \pi_{\text{\scriptsize aug}}(U; \gamma, \delta, \alpha)\} \{ f^\T(U), \psi^\T_\theta(U;\alpha)\}^\T.
$$
Assume that there exists a neighborhood $N_0$ of $\theta_0$ such that the following holds for any $\theta\in N_0$: $(\tilde\gamma, \tilde\delta)$ converges to a constant $(\gamma^\dag, \delta^*) = \{\gamma^\dag(\theta), \delta^*(\theta)\}$ such that
\begin{align*}
\left( \begin{array} {c}
 \tilde \gamma - \gamma^\dag \\
 \tilde\delta - \delta^*
 \end{array} \right) = {V^\dag}^{-1} \, \tilde E \left\{ s^\dag(T,U; \gamma^\dag, \delta^*, \hat\alpha) \right\} + o_p(n^{-1/2}),
\end{align*}
where
$E\{s^\dag(T,U; \gamma^\dag, \delta^*, \alpha^*)\}=0$, and the matrix
$V^\dag= - E\{ \partial s^\dag(T,U; \gamma, \delta, \alpha^*)/ $ $\partial (\gamma^\T, \delta^\T)\} |_{(\gamma,\delta)=(\gamma^\dag,\delta^*)}$ is nonsingular.
Write $\pi^\dag(U) = \pi_{\text{\scriptsize aug}}(U; \gamma^\dag, \delta^*, \alpha^*)$.
If model (\ref{TS:PS}) is correctly specified, then $(\gamma^\dag,\delta^*)=(\gamma^*,0)$, $\pi^\dag(U) = \pi(U)$,
$V ^\dag= \var \{ s^\dag(T,U;$ $ \gamma^*, 0, \alpha^* )  \}$,
and the asymptotic expansion for $(\tilde\gamma,\tilde\delta)$ reduces to
\begin{align*}
\left( \begin{array} {c}
 \tilde \gamma - \gamma^* \\
 \tilde\delta
 \end{array} \right) = {V^\dag}^{-1} \, \tilde E \left\{ s^\dag_{(\gamma^*,0)}(T,U) \right\} + o_p(n^{-1/2}),
\end{align*}
where $s^\dag_{(\gamma^*, 0)}(T,U) = s^\dag(T,U; \gamma^*, 0, \alpha^*)$.
\end{enumerate}

In addition, we assume that the following regularity conditions hold (e.g., \citealtappend{Robins1994}, Appendix B).
\begin{enumerate}
\item[(C4)] Assumptions 1--2 in \citeappend{Newey_Smith2004} hold for the vector of estimating functions $T \Phi(X,U;\theta)$.

\item[(C5)] There exists a constant $\epsilon>0$ such that $0<\pi^*(u) \le 1-\epsilon$ for all $u$.

\item[(C6)] There exists a constant $\epsilon>0$ such that $0<\pi_{\text{\scriptsize aug}}\{u; \gamma^\dag(\theta), \delta^*(\theta), \alpha^*\} \le 1-\epsilon$ for all $u$ and $\theta\in N_0$.

\item[(C7)] There exists a neighborhood $N_{1}$ of $\alpha^*$ such that $E\{ \sup_{\theta \in N_0, \alpha \in N_{1} } \| \partial \psi_\theta(U;\alpha)/ \partial \alpha \|^2 \} $ $  < \infty$,
where $\| A\| = (\sum_{ij} A_{ij}^2)^{1/2}$ for any matrix with element $A_{ij}$.

\item[(C8)]  There exists a neighborhood $N_2$ of $\gamma^*$ such that $E\{ \sup_{\gamma \in N_2 } \| \partial \pi(U;\gamma)/ \partial \gamma \|^2\}  < \infty$
and $E\{ \sup_{\gamma \in N_2 } \| \partial^2 \pi(U;\gamma)/ \partial \gamma\partial\gamma^\T \|^2 \}  < \infty$.

\item[(C9)]  There exists a neighborhood $N_{3,\theta}$ of $\{\gamma^\dag(\theta),\delta^*(\theta),\alpha^*\}$
such that $E\{ \sup_{ \theta\in N_0, \phi \in N_3 } $ $\| \partial \pi_{\text{\scriptsize aug}}(U;\phi)/ \partial\phi \|^2 \}  < \infty$
and $E\{ \sup_{\theta\in N_0, \phi \in N_3 } \| \partial^2 \pi_{\text{\scriptsize aug}}(U;\phi)/ \partial \phi\partial\phi^\T \|^2 \}  < \infty$,
where $\phi=(\gamma^\T,\delta^\T, \alpha^\T)^\T$ and $N_3 = \cup_{\theta\in N_0} N_{3,\theta}$.
\end{enumerate}

It can be shown that $\hat\theta_{\text{\scriptsize OR}}$ is consistent if OR model (\ref{general_OR_set}) is correctly specified,
and $\hat\theta_{\text{\scriptsize IPW}}$ is consistent if PS model (\ref{TS:PS}) is correctly specified.
These assumptions can also be used to justify various Taylor expansions in the following sections.

\subsection{Proof of Proposition \ref{prop2}}

For convenience, write $\Phi(\theta)=\Phi(X,U;\theta)$, $\hat{\pi}=\hat{\pi}(U)$, $\pi^*=\pi^*(U)$, $\pi=\pi(U)$, $\hat\psi_\theta=\hat\psi_\theta(U)$, $\psi^*_\theta = \psi^*_\theta(U)$,
and $\psi_\theta = \psi_\theta(U)$.

First, we show the double robustness of $\hat{\theta}_{\text{\scriptsize NP}}$.
If PS model (\ref{TS:PS}) is correctly specified, then $\pi^*=\pi$ and
$$
\tilde{E}\left\{\left( \frac{1-T}{1-\hat{\pi}}\hat{\pi} -T \right)\hat{\psi}_\theta \right\}=
\tilde{E}\left\{ \left( \frac{1-T}{1-\pi}\pi -T \right) {\psi}^*_\theta \right\}+O_p(n^{-1/2})=O_p(n^{-1/2}),
$$
and hence the left hand side of Eq.~(\ref{est_NP}) is
$$
\tilde{E}\left\{\frac{1-T}{1-\hat{\pi}}\hat{\pi}\Phi(\theta)\right\} +O_p(n^{-1/2}) .
$$
This implies that $\hat{\theta}_{\text{\scriptsize NP}}$ is a consistent estimator of $\theta_0$ similarly as $\hat{\theta}_{\text{\scriptsize IPW}}$.
If OR model (\ref{general_OR_set}) is correctly specified, then $\psi^*_\theta= \psi_\theta$ and
\begin{align*}
& \tilde{E}\left[ \frac{1-T}{1-\hat{\pi}} \hat\pi \{\Phi(\theta)- \hat \psi_\theta \} \right] = \tilde{E}\left[ \frac{1-T}{1-\pi^*} \pi^* \{\Phi(\theta)- \psi_\theta \} \right] = O_p(n^{-1/2}) ,
\end{align*}
and hence the left hand side of Eq.~(\ref{est_NP}) is
\begin{align*}
&  \tilde{E} (T \hat \psi_\theta ) +O_p(n^{-1/2}) .
\end{align*}
This implies that $\hat{\theta}_{\text{\scriptsize NP}}$ is a consistent estimator of $\theta_0$ similarly as $\hat{\theta}_{\text{\scriptsize OR}}$.

Second, we prove the local nonparametric efficiency of $\hat{\theta}_{\text{\scriptsize NP}}$.
If OR model (\ref{general_OR_set}) and PS model (\ref{TS:PS}) are correctly specified, then by Slutsky Theorem, the left hand side of Eq. (\ref{est_NP}) is
\begin{align*}
& \tilde{E}\left\{\frac{1-T}{1-\hat{\pi}}\hat{\pi}\Phi(\hat{\theta}_{\text{\scriptsize NP}})
- \left( \frac{1-T}{1-\hat{\pi}}\hat{\pi}-T \right) \psi_{\theta_0} \right\}+o_p(n^{-1/2}) \\
& = \tilde{E}\left[\frac{1-T}{1-\pi} \pi\big\{ \Phi(\hat \theta_{\text{\scriptsize NP}})-\psi_{\theta_0} \big\} + T\psi_{\theta_0} \right]+o_p(n^{-1/2}).
\end{align*}
By a Taylor expansion for $\hat{\theta}_{\text{\scriptsize NP}}$ about $\theta_0$, we have
\begin{align*}
\tilde E\left\{\frac{1-T}{1- {\pi}} {\pi}\frac{\partial\Phi(\theta_0)}{\partial\theta^\T}\right\}(\hat{\theta}_{\text{\scriptsize NP}}-\theta_0)
=- \tilde{E}\left[ \frac{1-T}{1-{\pi}}{\pi}\big\{ \Phi(\theta_0)-\psi\big\} + T\psi_{\theta_0} \right] +o_p(n^{-1/2}),
\end{align*}
and hence
\begin{align} \label{app-eq1}
\hat{\theta}_{\text{\scriptsize NP}}-\theta_0
&=- \Gamma_{\theta_0}^{-1} \times
\frac{1}{q}\tilde{E}\left[ \frac{1-T}{1-\pi}\pi\big\{\Phi(\theta_0)-\psi_{\theta_0} \big\} + T\psi_{\theta_0} \right] +o_p(n^{-1/2}).
\end{align}
Therefore, $\hat{\theta}_{\text{\scriptsize NP}}$ achieves the nonparametric variance bound $V_{\text{\scriptsize NP}}$ when both PS model and OR model are correctly specified.

Third, we prove the local semiparametric efficiency of $\hat{\theta}_{\text{\scriptsize SP}}$.
By similar arguments as in the derivation of Eq.~(\ref{app-eq1}), when OR model (\ref{general_OR_set}) and PS model (\ref{TS:PS}) are correctly specified,
we have
\begin{align} \label{app-eq2}
\hat{\theta}_{\text{\scriptsize SP}}-\theta_0
&=- \Gamma_{\theta_0}^{-1} \times
\frac{1}{q}\tilde{E}\left[ \frac{1-T}{1-\pi}\pi\big\{\Phi(\theta_0)-\psi_{\theta_0} \big\} + \hat \pi \psi_{\theta_0} \right] +o_p(n^{-1/2}).
\end{align}
By a Taylor expansion for $\hat\gamma$ about $\gamma$, we have
\begin{align*}
\tilde{E}(\hat\pi \psi_{\theta_0} )
=&\tilde{E}(\pi \psi_{\theta_0} )+E\left\{\psi_{\theta_0} \frac{\partial\pi(U;\gamma)}{\partial\gamma^{\T}} \right\}(\hat{\gamma}-\gamma)+o_p(n^{-1/2})\\
=&\tilde{E}(\pi \psi_{\theta_0} )+\mbox{Proj} \big\{ (T-\pi)\psi_{\theta_0} |S_{\gamma} \big\}+o_p(n^{-1/2}).
\end{align*}
Combining the preceding two expansions shows that $\hat{\theta}_{\text{\scriptsize SP}}$ achieves the semiparametric variance bound $V_{\text{\scriptsize SP}}$ when both OR and PS model are correctly specified. $\Box$

\subsection{Proof of Proposition \ref{prop-reg}}

First, it is straightforward to show that $\tilde\beta(\theta) = \beta^*(\theta) + o_p(1)$, where
$\beta^*(\theta) = E^{-1} (\xi^* {\zeta^*}^\T)$ $ E\{\xi^* \tau^*_{\text{\scriptsize init}}(\theta)\}$
and $\tau^*_{\text{\scriptsize init}}(\theta)$, $\xi^*$, $\zeta^*$, and $h^*(U)$ are defined as $\tilde{\tau}_{\text{\scriptsize init}}(\theta)$, $\tilde\xi$, $\tilde\zeta$, and $\tilde h(U)$ respectively
but with $\pi^\dag(U)$ and $\psi^*_\theta(U)$ in place of $\tilde \pi(U)$ and $\hat \psi_\theta(U)$ throughout.

Second, we show the local nonparametric efficiency and double robustness of $\tilde{\theta}_{\text{\scriptsize reg}}$.
By the discussion in Section \ref{sec:cal-reg}, it suffices to show that if OR model (\ref{general_OR_set}) is correctly specified, then
$\tilde\theta_{\text{\scriptsize reg}}$ is asymptotically equivalent, up to $o_p(n^{-1/2})$, to the solution of Eq.~(\ref{reason_DR}).
By construction, $\tilde{\pi}(U)\hat{\psi}_\theta (U)$ is a linear combination of variables in $\tilde{h}(U)/\tilde{\pi}(U)$, i.e.,
$\tilde{\pi}(U)\hat{\psi}_\theta (U)=c^\T(\theta) \tilde{h}(U)/\tilde{\pi}(U)$ for some constant vector $c(\theta)$.
Then $\pi^\dag(U)\psi^*_\theta (U)=c^\T(\theta) h^*(U)/\pi^{\dag}(U)$ also holds for the same vector $c(\theta)$. If OR model (\ref{general_OR_set}) is correctly specified,
then $\psi^*_\theta (U)=\psi_\theta (U)$
and $\pi^\dag(U)\psi(U)=c^\T h^*(U)/\pi^{\dag}(U)$. By direct calculation, we have
$$
\beta^*(\theta) =E^{-1}\left\{\xi^*\frac{1-T}{1-\pi^{\dag}(U)}\frac{h^{*\T}(U)}{\pi^{\dag}(U)}\right\}E\left\{\xi^*\frac{1-T}{1-\pi^{\dag}(U)}
\pi^{\dag}(U)\psi_\theta (U)\right\}=c( \theta) ,
$$
and hence the left hand side of Eq.~(\ref{est_reg}) is asymptotically equivalent, to the first order, to the left hand side of Eq.~(\ref{reason_DR}).

Third, we show the intrinsic efficiency of  $\tilde{\theta}_{\text{\scriptsize reg}}$  among the class of estimators, denoted by $\tilde{\theta}(b)$, that are solutions to (\ref{theta_class}).
By similar arguments as in the derivation of Eqs.~(\ref{app-eq1}) and (\ref{app-eq2}), when PS model (\ref{TS:PS}) is correctly specified, we have
\begin{align*}
\tilde{\theta}(b)-\theta_0=- \Gamma_{\theta_0}^{-1}
\times \frac{1}{q}\tilde{E}\left\{ \frac{1-T}{1-\tilde{\pi}}\tilde{\pi}\Phi(\theta_0)-b^\T \left(\frac{1-T}{1-\tilde{\pi}}-1\right) \frac{h^*}
{\pi}\right\} +o_p(n^{-1/2}),
\end{align*}
where $\theta$ is fixed at the true $\theta_0$ in $h^*$, $\xi^*$, and $\zeta^*$.
By applying Lemma 1 in \citetappend{Shu2015} with $\hat{\pi}$ replaced by $\tilde{\pi}$, $Y$ by $\Phi(\theta_0)$, and $h$ by $b^\T h^*/\pi$, we have
\begin{align*}
 \tilde{\theta}(b)-\theta_0 &= - \Gamma_{\theta_0}^{-1}\times \frac{1}{q}\tilde{E}\Big[\tau_{\text{\scriptsize init}}^*(\theta_0)-b^{\T}\xi^*
-\mbox{Proj}\big\{\tau_{\text{\scriptsize init}}^*(\theta_0)-b^{\T}\xi^*|S_{(\gamma^*,0)}^\dag\big\} \\
& \quad + \mbox{Proj} \big\{(T-\pi)\psi_{\theta_0} |S_{(\gamma^*,0)}^\dag\big\}\Big]  +o_p(n^{-1/2}).
\end{align*}
Inside $\tilde{E}()$ above, the first term $\tau_{\text{\scriptsize init}}^*(\theta_0)-b^{\T}\xi^*
-\mbox{Proj} \big\{\tau_{\text{\scriptsize init}}^*(\theta_0)-b^{\T}\xi^*|S_{(\gamma^*,0)}^\dag\big\}$ is uncorrelated with the second term $\mbox{Proj}\big\{(T-\pi)\psi_{\theta_0} |S_{(\gamma^*,0)}^\dag\big\}$, which is independent of $b$. Moreover, the first term can be expressed as $\tau_{\text{\scriptsize init}}^*(\theta_0)-a^{\T}\xi^*$ for some constant vector $a$, because each variable in $S_{(\gamma^*,0)}^\dag$ is a linear combination of variables in $\xi^*$ by construction. By combining these two facts, the asymptotic variance of $\tilde{\theta}(b)$ is minimized when $a$ is equal to
\begin{align*}
 \var^{-1}(\xi^*) \cov  \left\{ \xi^*, \, \tau_{\text{\scriptsize init}}^*(\theta_0)\right\} = E^{-1} ( \xi^* {\zeta^*}^\T)  E\{\xi^*\tau_{\text{\scriptsize init}}^*(\theta_0)\} = \beta^* .
\end{align*}
But to make $a$ equal to $\beta^*$, it suffices to set $b = \beta^*(\theta_0)$, because $\tau_{\text{\scriptsize init}}^*(\theta_0) - {\beta^*}^\T \xi^*$ is uncorrelated with $S^\dag_{(\gamma^*,0)}$
and hence $\mbox{Proj}\{ \tau_{\text{\scriptsize init}}^*(\theta_0) - {\beta^*}^\T \xi^*| S^\dag_{(\gamma^*,0)}\}  = 0$.
If PS model (\ref{TS:PS}) is correctly specified, then $\tilde\theta_{\text{\scriptsize reg}} = \tilde \theta( \beta^*) + o_p(n^{-1/2})$.
Therefore, $\tilde\theta_{\text{\scriptsize reg}} $ is intrinsically efficient among the class of estimators $\tilde \theta(b)$. $\Box$

\subsection{Proof of Corollary~\ref{cor-reg}}

The comparison follows from Proposition~\ref{prop-reg} directly for $\tilde{\theta}_{\text{\scriptsize IPW}}$, which falls in the class (\ref{theta_class}) with $b=0$.
The estimating equation (\ref{TS:tilde_NP}) for $\tilde{\theta}_{\text{\scriptsize NP}}$ can be rewritten as
\begin{align*}
0 & = \tilde{E}\left[ \tilde \tau_{\text{\scriptsize init}}(\theta) -  \frac{1-T}{1-\tilde{\pi}(U)}\tilde{\pi}(U) \hat{\psi}_\theta (U) + T\hat{\psi}_\theta (U)\right]\\
& =  \tilde{E}\left[ \tilde \tau_{\text{\scriptsize init}}(\theta) - \left\{ \frac{1-T}{1-\tilde{\pi}(U)}-1\right\}\tilde{\pi}(U) \hat{\psi}_\theta (U) + \{ T-\tilde{\pi}(U)\} \hat{\psi}_\theta (U)\right],
\end{align*}
which is of the form (\ref{theta_class}) for some suitable $b$, because both $[(1-T)/\{1-\tilde{\pi}(U)\}-1]\tilde{\pi}(U)\hat{\psi}_\theta (U)$ and $\{T-\tilde{\pi}(U)\}\hat{\psi}_\theta (U)$ are included in $\tilde{\xi}$.
For the estimator $\tilde{\theta}_{\text{\scriptsize AST}}$, if PS model (\ref{TS:PS}) is correctly specified, then $\tilde \chi$ converges to 0 in probability and hence
\begin{align*} \tilde \chi & =- \tilde E^{-1} \left[\frac{ (1-T) \tilde\pi(U) \hat \psi_\theta(U) }{ \{1-\tilde \pi(U)\}^2}
\frac{\partial\pi_{\text{\scriptsize aug}}}{\partial \chi^\T} (U;\tilde\gamma,\tilde\delta+\chi,\hat\alpha) \big|_{\chi=0} \right] \times  \\
& \quad \tilde E \left[ \left\{\frac{1-T}{ 1-\tilde \pi(U)}-1 \right\} \tilde \pi(U) \hat\psi_\theta(U)\right] + o_p(n^{-1/2})
\end{align*}
by a Taylor expansion of $\tilde E \big[\{ (1-T) \tilde w^{-1}_{\text{\scriptsize AST}}(U)-1\} \tilde \pi(U) \hat \psi_\theta(U) \big]=0$.
Moreover, a Taylor expansion of the estimating equation for $\tilde{\theta}_{\text{\scriptsize AST}}$ gives
\begin{align*}
0 &= 
\tilde E \left\{ \frac{(1-T) \tilde \pi(U)}{1 - \tilde \pi(U)} \Phi(X,U;\theta) \right\}+ \\
& \quad \tilde E^{-1} \left[\frac{ (1-T) \tilde\pi(U) \Phi(X,U;\theta) }{ \{1-\tilde \pi(U)\}^2}
\frac{\partial\pi_{\text{\scriptsize aug}}}{\partial \chi^\T} (U;\tilde\gamma,\tilde\delta+\chi,\hat\alpha) \big|_{\chi=0}\right] \tilde \chi + o_p(n^{-1/2}) .
\end{align*}
Therefore, if PS model (\ref{TS:PS}) is correctly specified, then $\tilde{\theta}_{\text{\scriptsize AST}}$ is, up to $o_p(n^{-1/2})$, a solution to estimating equations of the form (\ref{theta_class}) for some suitable $b$.
The conclusion then follows from  Proposition~\ref{prop-reg}. $\Box$

\subsection{Derivation of empirical likelihood estimates}

By standard calculation (\citealtappend{Qinlawless1994}), the empirical likelihood estimate of $p_i$ subject to constraints (\ref{emp-lik-constraints}) is
\begin{align*}
\hat p_i = \frac{n^{-1}}{1 + \hat\lambda^\T \tilde \xi_{i}},
\end{align*}
where $\hat\lambda$ is a maximizer of the function
\begin{align*}
\ell_{\text{\scriptsize EL}}(\lambda) = \frac{1}{n} \sum_{i=1}^n \log \left( 1 + \lambda^\T \tilde \xi_{i} \right).
\end{align*}
Write $\tilde\pi_i = \tilde \pi(U_i)$, $\tilde h_i = \tilde h(U_i)$, and $\omega_i = \omega(U_i;\lambda)=\tilde{\pi}_i+{\lambda}^{\T}\tilde{h}_i$ for $i=1,\ldots,n$.
By direct calculation, $\ell_{\text{\scriptsize EL}}(\lambda)$ can be reexpressed as
\begin{align*}
& \ell_{\text{\scriptsize EL}}(\lambda)  = \frac{1}{n}\sum_{i=1}^n \log \left\{  1+ \lambda^\T \frac{T_i-\tilde{\pi}_i}{\tilde{\pi}_i(1-\tilde{\pi}_i)}\tilde{h}_i  \right\} \\
& = \frac{1}{n}\sum_{i=1}^n \left\{ T_i \log\left(  1+ \lambda^\T \frac{\tilde{h}_i }{\tilde{\pi}_i} \right) + (1- T_i ) \log \left(  1- \lambda^\T \frac{\tilde{h}_i }{1-\tilde{\pi}_i} \right) \right\} \\
& = \frac{1}{n}\sum_{i=1}^n \left\{ T_i\log\omega_i+ (1-T_i)\log(1-\omega_i) \right\} -\frac{1}{n}\sum_{i=1}^n \left\{ T_i\log\tilde{\pi}_i + (1-T_i)\log(1-\tilde{\pi}_i) \right\},
\end{align*}
which equals $\ell(\lambda)$ up to an additive constant.
Therefore, $\hat{\lambda}$ is a maximizer of $l(\lambda)=\tilde{E}[T\log\omega(U;\lambda)+(1-T)\log\{1-\omega(U;\lambda)\}]$.

Write $\Phi_i(\theta)=\Phi(X_i,U_i; \theta)$. Eq.~(\ref{lik_ndr}) can be reexpressed as Eq.~(\ref{lik1}) because
\begin{align*}
0 &=\sum_{i=1}^n \hat{p}_i\left\{ \frac{1-T_i}{1-\tilde{\pi}_i}\tilde{\pi}_i\Phi_i(\theta)\right\}
=\frac{1}{n}\sum_{i=1}^n\frac{1-T_i}{1+\hat{\lambda}^\T\tilde{\xi}_i}\frac{\tilde{\pi}_i}{1-\tilde{\pi}_i}\Phi_i(\theta) \\
& =\frac{1}{n}\sum_{i=1}^n\frac{1-T_i}{1-\hat\lambda^\T\frac{\tilde h_i}{1-\tilde \pi_i}}\frac{\tilde{\pi}_i}{1-\tilde{\pi}_i}\Phi_i(\theta)
=\frac{1}{n} \sum_{i=1}^n \frac{1-T_i}{1-\hat\omega_i}\tilde \pi_i \Phi_i(\theta) ,
\end{align*}
where $\hat\omega_i = \omega(U_i; \hat\lambda)$ for $i=1,\ldots,n$. $\Box$

\subsection{Proof of Proposition \ref{prop-lik}}

We need only to show that if PS model (\ref{TS:PS}) is correctly specified, then
$\tilde\theta_{\text{\scriptsize lik}}$ is asymptotically equivalent, to the first order, to $\tilde\theta_{\text{\scriptsize reg}}$.
If PS model (\ref{TS:PS}) is correctly specified, the left hand side of Eq.~(\ref{eq_general_lik}) can be approximated as
\begin{align*}
\tilde{E}\left[ \frac{(1-T)\tilde{\pi}(U)\Phi(X,U;\theta)}{1-\omega(U ;\tilde\lambda)} \right] & =\tilde{E}\left[ \frac{(1-T)\tilde{\pi}(U)\Phi(X,U;\theta)}{1-\omega(U ;\hat\lambda)} \right] + o_p(n^{-1/2}),
\end{align*}
by a Taylor expansion for $\tilde \lambda$ about $\hat\lambda$ and the fact that $\tilde E ( [ (1-T)/\{1-\omega(U;\hat\lambda)\}-1 ] \tilde v(U) ) = o_p(n^{-1/2})$,
similarly as in the asymptotic expansion of the calibrated likelihood estimator in \citetappend{Tan2010}.
Moreover, if PS model (\ref{TS:PS}) is correctly specified, then $\hat\lambda$ converges to $0$ in probability and
\begin{align*}
\tilde{E}\left[ \frac{(1-T)\tilde{\pi}(U )\Phi(X,U;\theta)}{1-\omega(U ;\hat\lambda)} \right]& =\tilde{E}\left\{\tilde{\tau}_{\text{\scriptsize init}}(\theta) - {\beta^*}^\T(\theta) \tilde{\xi }\right\} + o_p(n^{-1/2}).
\end{align*}
by a Taylor expansion for $\hat\lambda$ about $0$, similarly as in the asymptotic expansion of the non-calibrated likelihood estimator in \citetappend{Tan2010}.
The desired result for $\tilde\theta_{\text{\scriptsize lik}}$ then follows from the preceding the expansions.
$\Box$

\newpage
\section{Additional simulation results}

We present additional simulation results in the setup of Section \ref{sec:simulation}, but
with Eq.~(\ref{x_primary}) modified to
\begin{align*}
X = 0.8 Z_0 + 0.6 Z_1 -0.5 Z_2+ e,
\end{align*}
or modified to
\begin{align*}
X = 0.6 Z_0 + 0.6 Z_1 -0.5 Z_2+ e.
\end{align*}
That is, the coefficient of the instrument $Z_0$ is set to $0.8$ or $0.6$.
The correlation between $X$ and the instrument $Z_0$ is
$.62$, $.53$, or $.43$ respectively in the case of IV coefficient 1, $0.8$, or $0.6$,
corresponding to an increasingly weaker instrument.

\begin{figure}
\caption{\small Boxplots of estimates of $\beta$ relative to $0.5$ ($n_1=5000$, $n_0=500$, IV coef $=0.8$)} \vspace{.1in}
\label{plot:TSIVboxplots-coef08}
\begin{tabular}{c}
\includegraphics[width=5.7in, height=4in]{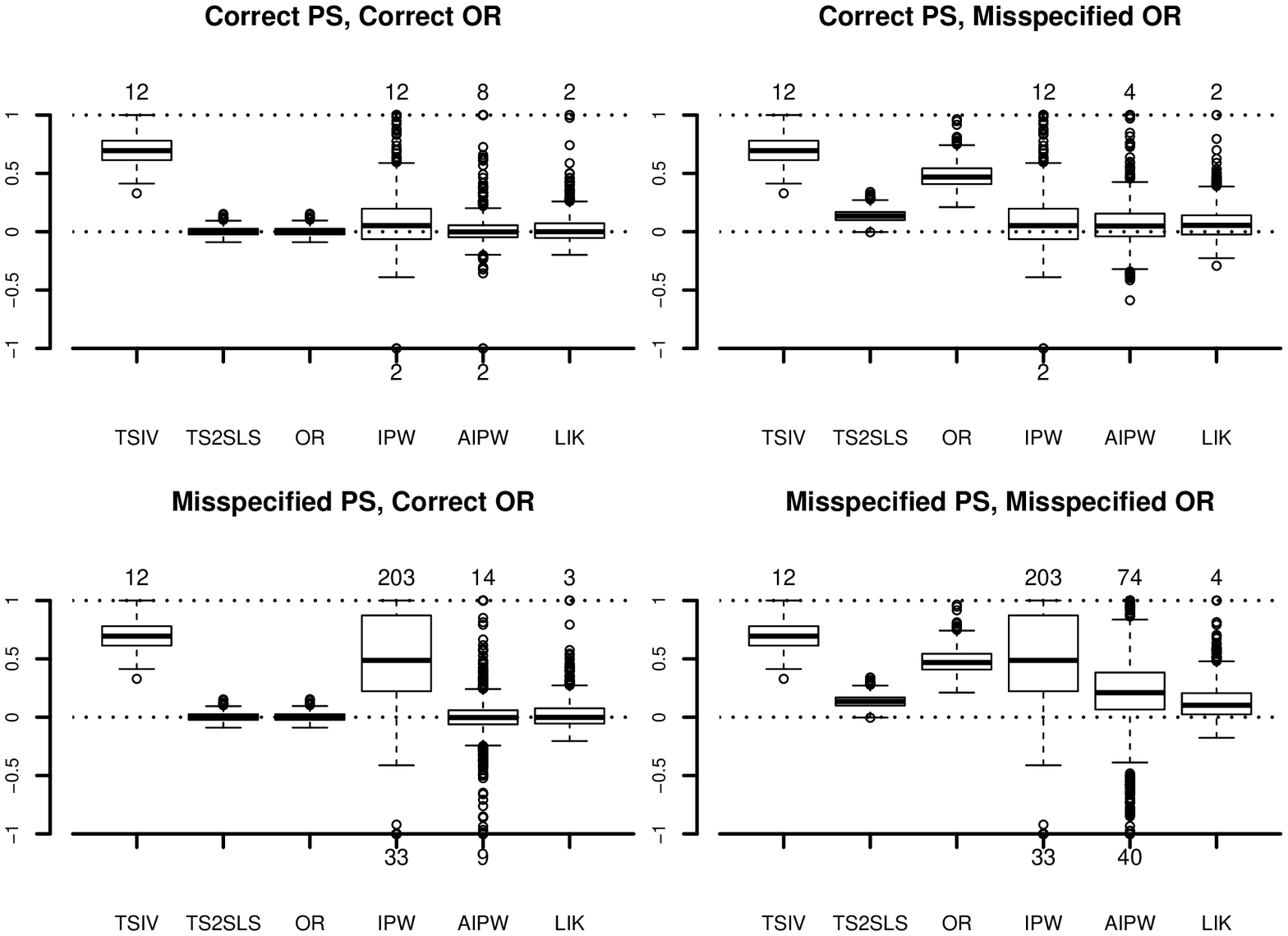}
\end{tabular}
\end{figure}

\begin{figure}
\caption{\small Boxplots of estimates of $\beta$ relative to $0.5$ ($n_1=5000$, $n_0=500$, IV coef $=0.6$)} \vspace{.1in}
\label{plot:TSIVboxplots-coef06}
\begin{tabular}{c}
\includegraphics[width=5.7in, height=4in]{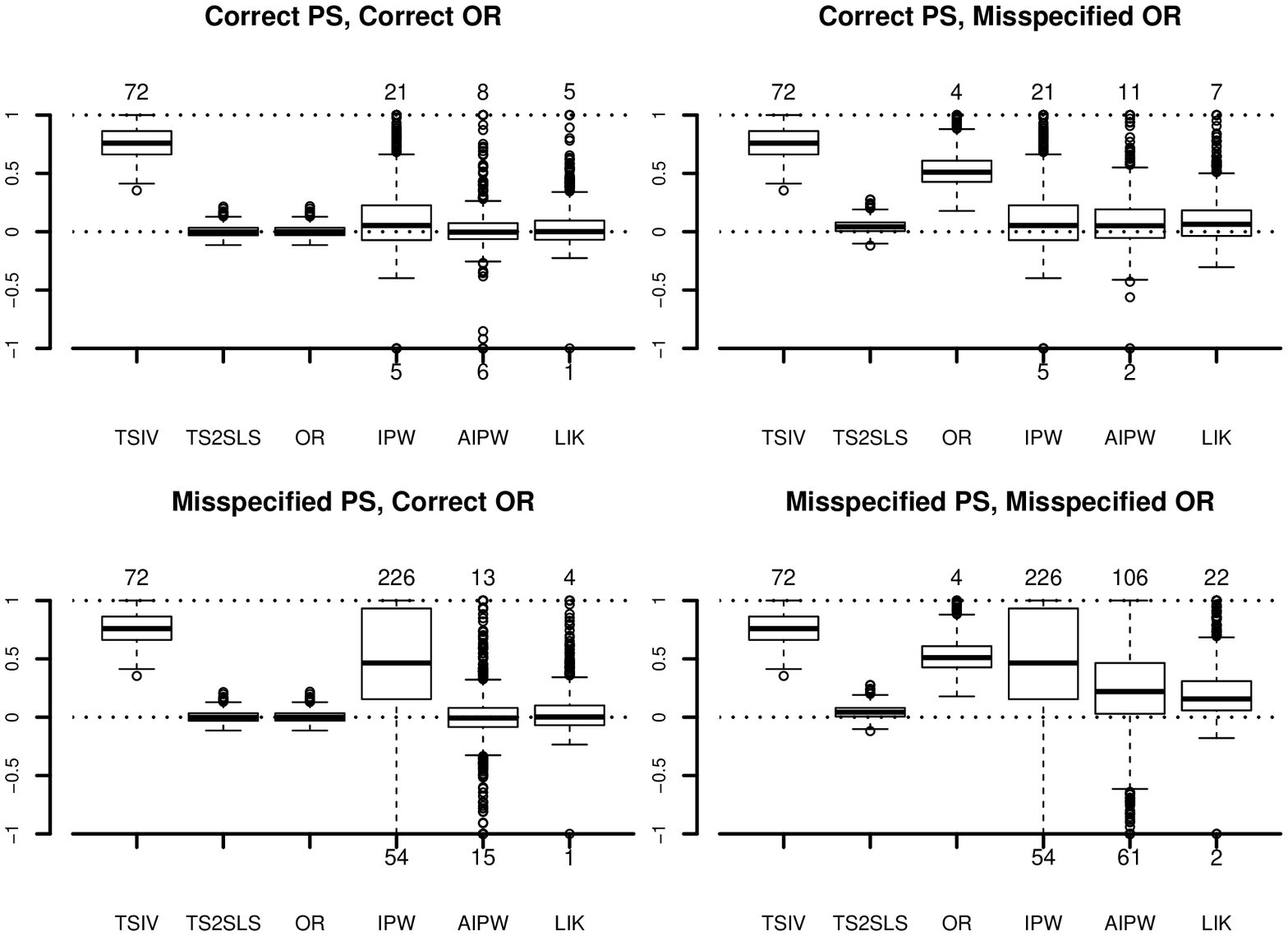}
\end{tabular}
\end{figure}

Table~\ref{simu:comp-coef08} and Figure~\ref{plot:TSIVboxplots-coef08} summarize the results in the case of IV coefficient $0.8$,
and Table~\ref{simu:comp-coef06} and Figure~\ref{plot:TSIVboxplots-coef06} summarize the results in the case of IV coefficient $0.6$,
similarly as Table~\ref{simu:comp} and Figure~\ref{plot:TSIVboxplots} in the case of IV coefficient $1$.
All estimators lead to larger standard errors when the coefficient (or strength) of the instrument decreases.
But the relative performances of these estimators in the case of IV coefficient $0.8$ or $0.6$ are similar to those in the case of  IV coefficient $1$.

\clearpage
\begin{table}
   \caption{Summary of estimates of $\beta$ ($n_1=5000$, $n_0=500$, IV coef $=0.8$)}
   \scriptsize
   \label{simu:comp-coef08}
   \begin{center}
   \begin{tabular}{lcccccc}
   \hline\hline
\noalign{\medskip}
                       & TSIV      & TS2SLS    & OR        & IPW & AIPW      & LIK       \\
\noalign{\medskip}\hline\noalign{\medskip}
Correct PS, Correct OR &0.70214	& 0.00177	& 0.00206	& 0.08849	& 0.07985	& 0.02405   \\
                       &(0.12170)	& (0.03621)	& (0.03629)	& (0.46412)	& (1.74026)	& (0.13319) \\
                       \noalign{\medskip}
Correct PS, Misspecified OR  &0.70214	& 0.13775	& 0.48176	& 0.08849	& 0.06736	& 0.07169   \\
                       &(0.12170)	& (0.05393)	& (0.10262)	& (0.46412)	& (0.20352)	& (0.14197) \\
                       \noalign{\medskip}
Misspecified PS, Correct OR   &0.70214	& 0.00177	& 0.00206	& -0.48397	& 0.01444	& 0.02720  \\
                       &(0.12170)	& (0.03621)	& (0.03629)	& (33.07874)	& (0.46850)	& (0.14907) \\
                       \noalign{\medskip}
Misspecified PS, Misspecified OR   &0.70214	& 0.13775	& 0.48176	& -0.48397	& 0.16885	& 0.13498   \\
                       &(0.12170)	& (0.05393)	& (0.10262)	& (33.07874)	& (2.12672)	& (0.19056)\\
\noalign{\medskip}\hline\hline
\end{tabular}
   \end{center}
   \vspace{-0.1in}
The estimators of $\beta$ are taken from $\hat\beta^\dag_{\text{\scriptsize TSIV}}$, $\hat\beta^\dag_{\text{\scriptsize TS2SLS}}$, and $\hat\beta^\dag$ in Eq.~(\ref{general_TS_est})
where $\hat\mu_3$ is set to $\hat\mu_{3,\text{OR}}$, $\hat\mu_{3,\text{IPW}}$, $\hat\mu_{3,\text{AIPW}}$, and $\tilde\mu_{3,\text{lik}}$ with $\tilde h_2$ removed from $\tilde h$.
Each cell gives the empirical bias (upper) and standard deviation (lower) of the point estimator, from 1000 Monte Carlo samples.
  \end{table}

\normalsize

\begin{table}
   \caption{Summary of estimates of $\beta$ ($n_1=5000$, $n_0=500$, IV coef $=0.6$)}
   \scriptsize
   \label{simu:comp-coef06}
   \begin{center}
   \begin{tabular}{lcccccc}
   \hline\hline
\noalign{\medskip}
                       & TSIV      & TS2SLS    & OR        & IPW & AIPW      & LIK       \\
\noalign{\medskip}\hline\noalign{\medskip}
Correct PS, Correct OR & 0.76775  & 0.00300	& 0.00353	& 0.03036	& -0.01200	& 0.02697  \\
                       & (0.14729)	& (0.04866)	& (0.04884) & 	(2.40077) & 	(3.51675) &	(0.40818) \\
                       \noalign{\medskip}
Correct PS, Misspecified OR   & 0.76775	& 0.04516	& 0.53005	& 0.03036	& 0.55899	& 0.10943   \\
                       & (0.14729)	& (0.05885)	& (0.14107) &	(2.40077)	& (18.90818) & (0.34001) \\
                       \noalign{\medskip}
Misspecified PS, Correct OR   & 0.76775	& 0.00300	& 0.00353	& -4.37817	& -0.00123	& 0.03718  \\
                       & (0.14729)	& (0.04866)	& (0.04884)	& (156.98569)	& (1.38405)	& (0.27388) \\
                       \noalign{\medskip}
Misspecified PS, Misspecified OR    & 0.76775	& 0.04516	& 0.53005	& -4.37817	& 0.63544	& 0.21258   \\
                       & (0.14729)	& (0.05885) & (0.14107)	& (156.98569) &	(9.17403)	& (0.68056) \\
\noalign{\medskip}\hline\hline
\end{tabular}
   \end{center}
   \vspace{-0.1in}
The estimators of $\beta$ are taken from $\hat\beta^\dag_{\text{\scriptsize TSIV}}$, $\hat\beta^\dag_{\text{\scriptsize TS2SLS}}$, and $\hat\beta^\dag$ in Eq.~(\ref{general_TS_est})
where $\hat\mu_3$ is set to $\hat\mu_{3,\text{OR}}$, $\hat\mu_{3,\text{IPW}}$, $\hat\mu_{3,\text{AIPW}}$, and $\tilde\mu_{3,\text{lik}}$ with $\tilde h_2$ removed from $\tilde h$.
Each cell gives the empirical bias (upper) and standard deviation (lower) of the point estimator, from 1000 Monte Carlo samples.
  \end{table}

\normalsize

\begin{figure}[h]
\caption{Scatterplots of $X$ versus transformed variables in the auxiliary data and boxplots of transformed variables between the two samples ($n_1=500$, $n_0=5000$)}
\label{scatterplot2}
\begin{tabular}{c}
\includegraphics[width=4in, height=5.7in, angle=270]{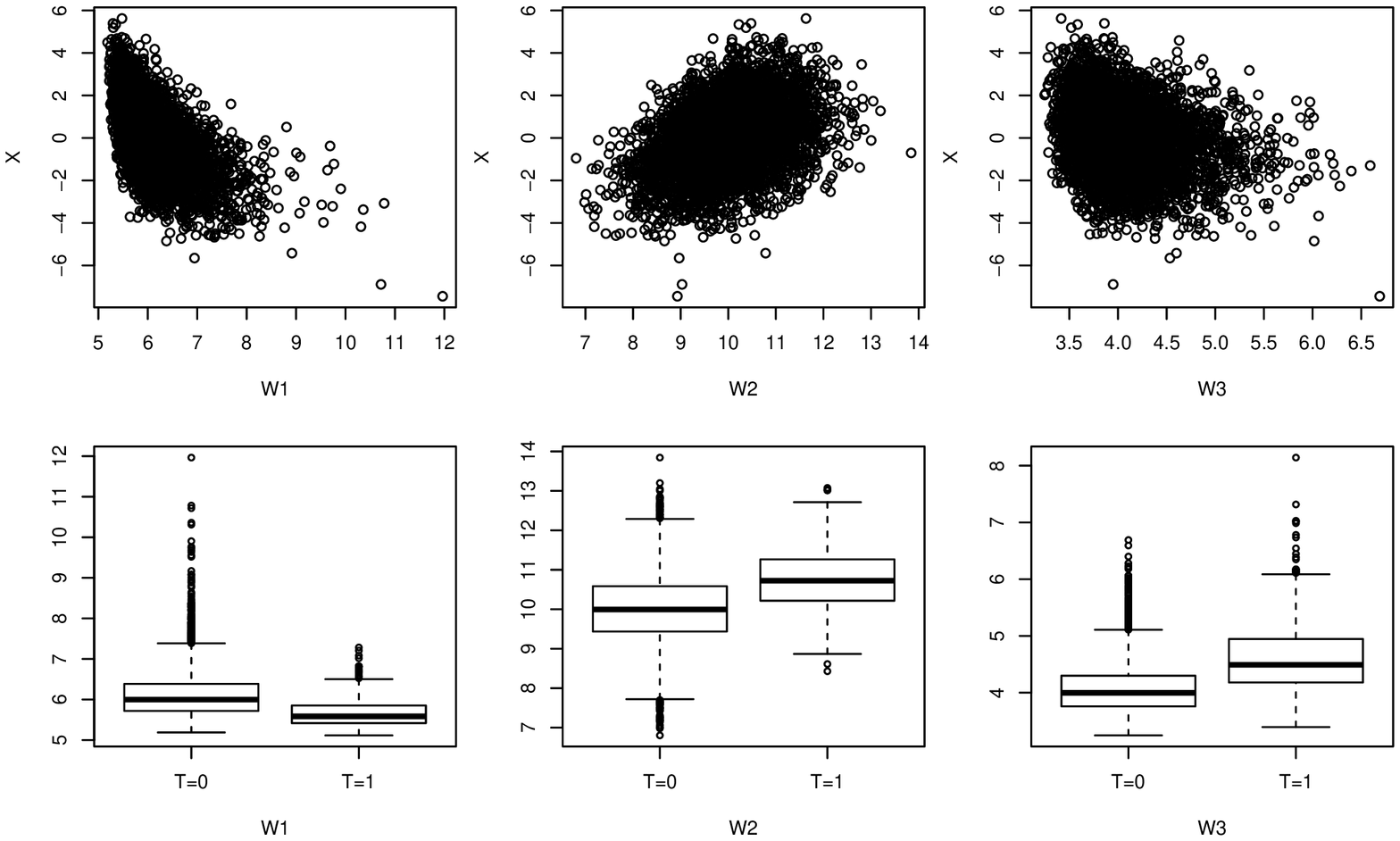}
\end{tabular}
\end{figure}

\begin{figure}
\caption{\small Boxplots of estimates of $\beta$ relative to $0.5$ ($n_1=500$, $n_0=5000$)} \vspace{.1in}
\label{plot:TSIVboxplots2}
\begin{tabular}{c}
\includegraphics[width=5.7in, height=4in]{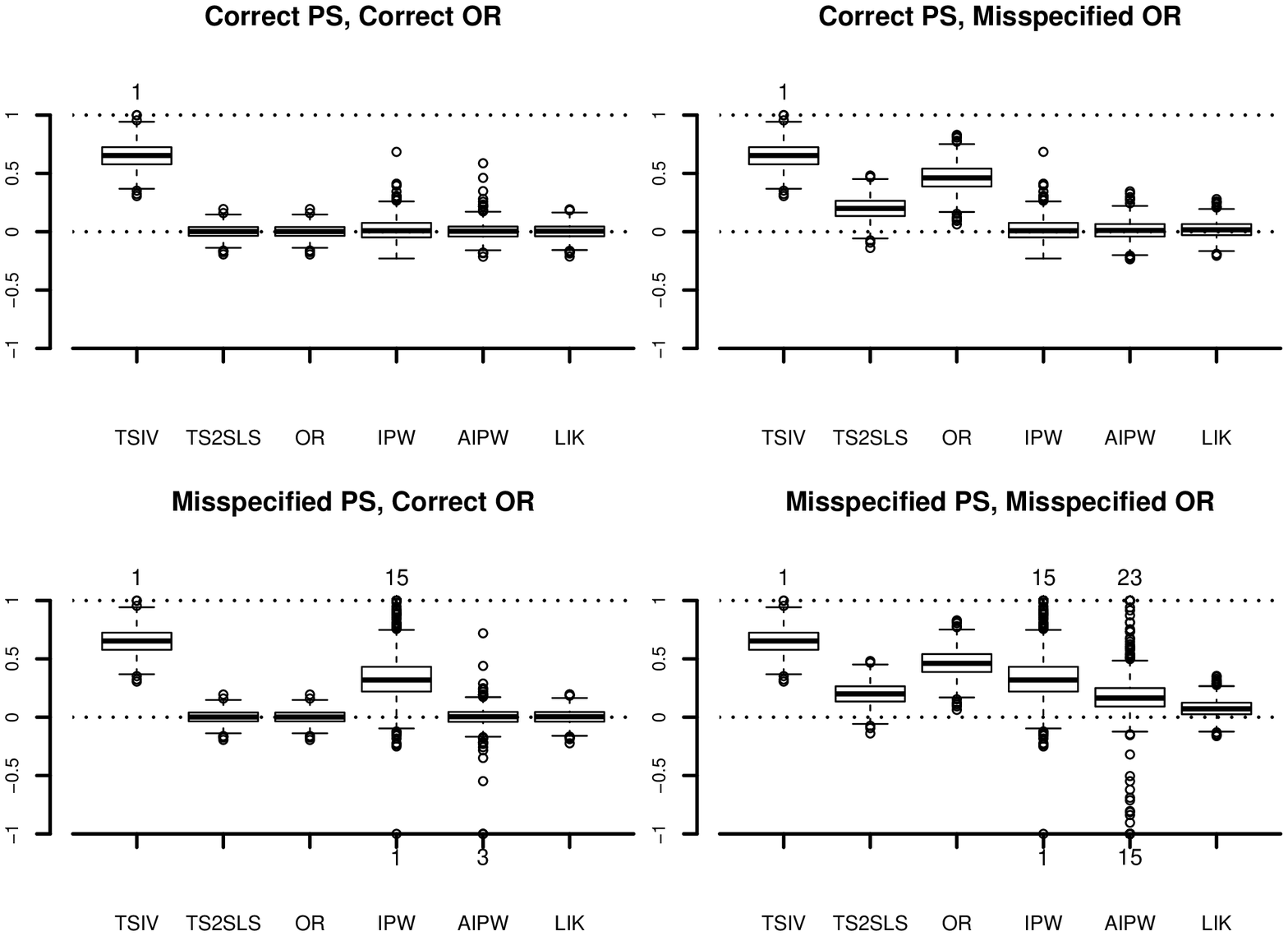}
\end{tabular}
\end{figure}

\clearpage
We present additional simulation results with the primary and auxiliary data sizes $n_1=5000$ and $n_0=500$.
The true propensity score is
\begin{align*}
P(T=1| U )=\text{expit} (-1.5 -\log10+ Z_0+ Z_1+ Z_2 ).
\end{align*}
Figures \ref{scatterplot2} and \ref{plot:TSIVboxplots2} and Table \ref{simu:comp2} correspond to Figures \ref{scatterplot} and \ref{plot:TSIVboxplots} and Table \ref{simu:comp} respectively.

The TS2SLS and OR estimators have larger variances, whereas the IPW, AIPW, and LIK  estimators have smaller variances, than in the case of $n_1=5000$ and $n_0=500$,
because the variability of IPW based estimators is affected mainly by the auxiliary data size.
Nevertheless, the relative performances of the estimates of $\beta$ are qualitatively the same as in the case of $n_1=5000$ and $n_0=500$.

\vspace{.2in}

\begin{table}
   \caption{Summary of estimates of $\beta$ ($n_1=500$, $n_0=5000$)}
   \scriptsize
   \label{simu:comp2}
   \begin{center}
   \begin{tabular}{lcccccc}
   \hline\hline
\noalign{\medskip}
                       & TSIV      & TS2SLS    & OR        & IPW & AIPW      & LIK       \\
\noalign{\medskip}\hline\noalign{\medskip}
Correct PS, Correct OR &0.65155	& 0.00235	& 0.00235	& 0.01568	& 0.00526	& 0.00410  \\
                       &(0.11001)	& (0.05461)	& (0.05462)	& (0.09699)	& (0.06781)	& (0.06109) \\
                       \noalign{\medskip}
Correct PS, Misspecified OR  &0.65155	& 0.19988	& 0.46434	& 0.01568	& 0.01274	& 0.01849  \\
                       & (0.11001)	& (0.09562)	& (0.11383)	& (0.09699)	& (0.08382)	& (0.07263) \\
                       \noalign{\medskip}
Misspecified PS, Correct OR   &0.65155	& 0.00235	& 0.00235	& 0.46519	& -0.03335	& 0.00441  \\
                       &(0.11001)	& (0.05461)	& (0.05462)	& (2.41001) & (0.97861)	& (0.06182) \\
                       \noalign{\medskip}
Misspecified PS, Misspecified OR   &0.65155	& 0.19988	& 0.46434	& 0.46519	& 0.19439	& 0.07495   \\
                       &(0.11001)  & (0.09562)	& (0.11383)	& (2.41001)	& (1.10682)	& (0.07773) \\
\noalign{\medskip}\hline\hline
\end{tabular}
   \end{center}
   \vspace{-0.1in}
The estimators of $\beta$ are taken from $\hat\beta^\dag_{\text{\scriptsize TSIV}}$, $\hat\beta^\dag_{\text{\scriptsize TS2SLS}}$, and $\hat\beta^\dag$ in Eq.~(\ref{general_TS_est})
where $\hat\mu_3$ is set to $\hat\mu_{3,\text{OR}}$, $\hat\mu_{3,\text{IPW}}$, $\hat\mu_{3,\text{AIPW}}$, and $\tilde\mu_{3,\text{lik}}$ with $\tilde h_2$ removed from $\tilde h$.
Each cell gives the empirical bias (upper) and standard deviation (lower) of the point estimator, from 1000 Monte Carlo samples.
 \end{table}

\normalsize

\newpage
\section{Additional results on housing projects}

\begin{figure}[h]
\caption{\small Boxplots of bootstrap estimates of $\beta$ (housing density)}  \vspace{-.1in}
\label{boxplots:housing2}
\begin{tabular}{c}
\includegraphics[width=5.7in, height=2in]{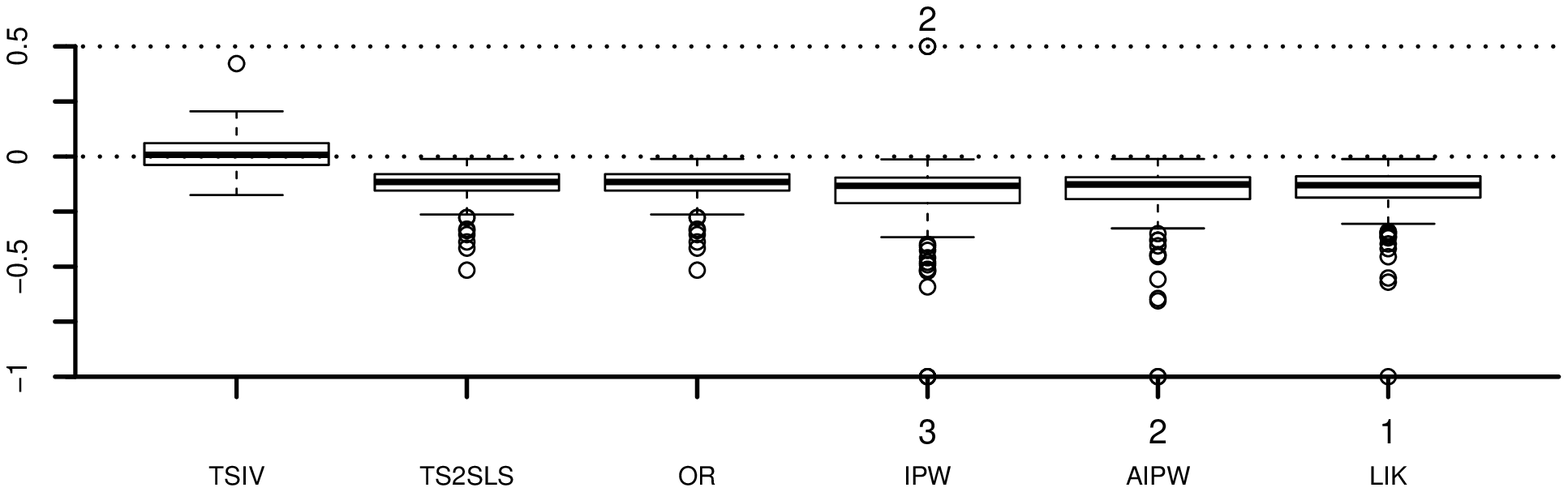}
\end{tabular}
\end{figure}

We present the results for reassessing the effect of project participation on the outcome $Y$, housing density,
which is defined as 1 if a family lived in high-density housing, that is, a building with over 50 units.
Table \ref{results of real data2} and Figure \ref{boxplots:housing2} correspond to Table \ref{results of real data} and Figure \ref{boxplots:housing} respectively.

The results from AIPW and LIK agree with that from TS2SLS, but are more robust than the latter to possible violations of modeling assumptions.
Moreover, the 95\% bootstrap confidence interval from LIK is much narrower than from AIPW.
All of these results suggest
that families in housing projects are less likely to live in high-density buildings.

\vspace{.2in}
\begin{table}[h]
   \caption{Estimates of the effect of project participation on housing density}  \vspace{-.1in}
   \scriptsize
   \label{results of real data2}
   \begin{center}
   \begin{tabular}{lcccccc}
  \hline\hline
\noalign{\medskip} &TSIV    & TS2SLS            & OR                & IPW               & AIPW     & LIK    \\
\noalign{\smallskip} \hline   \noalign{\smallskip}
Point &0.01108	& -0.11539	& -0.11539	& -0.14787	& -0.13618	& -0.13233      \\
SE &0.06838	& ---	& 0.04689	& 0.08427	& 0.07235	& 0.07020        \\
boot.SE &0.07551	& 0.07356	& 0.07356	& 4.46305	& 3.47457	& 0.12808           \\
boot.CI &(-0.119, 0.150)	& (-0.330, -0.024)	& (-0.330, -0.024)	& (-0.520, -0.024)	& (-0.457, -0.026)	& (-0.397, -0.026)\\
   \noalign{\smallskip}
         \hline\hline
\end{tabular}
   \end{center}
\vspace{-0.1in}
Each column gives the point estimate (upper), the analytical and bootstrap (boot) standard errors (middle), and 95\% bootstrap percentile confidence interval (lower) from 200 bootstrap samples.
For comparison, the TS2SLS estimate is reported as $-0.1154$, with analytical standard error $0.0468$, in \citetappend{Currie2000}.
   \end{table}

\clearpage

\setlength{\bibsep}{3pt}

\bibliographystyleappend{myapa}
\bibliographyappend{thesisnew2}

\end{document}